\documentclass[11pt,a4paper]{article}
\usepackage{jcappub}
\usepackage[utf8]{inputenc}
\pdfoutput=1 
\usepackage{graphicx}
\usepackage[figuresright]{rotating}
\usepackage{amssymb}
\usepackage{bm}
\usepackage{color}
\usepackage{placeins}
 \usepackage{textcomp}
\usepackage{ulem} 
\usepackage{booktabs} 
\usepackage{tablefootnote}
\usepackage{graphicx}
\usepackage{caption}
\usepackage{subcaption}
\usepackage{array}
\usepackage[shortlabels]{enumitem}
\usepackage{float}

\definecolor{dgreen}{rgb}{0,0.6,0.0}

\newcommand{\HH}{{\cal H}}

\newcommand{\al}{\alpha}

\newcommand{\de}{\delta}
\newcommand{\De}{\Delta}

\newcommand{\ga}{\gamma}

\newcommand{\La}{\Lambda}
\newcommand{\la}{\lambda}
\newcommand{\Om}{\Omega}

\newcommand{\si}{\sigma}

\newcommand{\ra}{\rightarrow}

\newcommand{\be}{\begin{equation}}
\newcommand{\ee}{\end{equation}}

\newcommand{\lsim}{\stackrel{<}{\sim}}
\newcommand{\bea}{\begin{eqnarray}}
\newcommand{\eea}{\end{eqnarray}}
\newcommand{\bean}{\begin{eqnarray*}}
\newcommand{\eean}{\end{eqnarray*}}

\newcommand{\id}{{\rm 1\kern -2.5pt I}} 
\newcommand{\bn}{{\mathbf n}}
\newcommand{\bv}{{\mathbf v}}
\newcommand{\bw}{{\mathbf w}}
\newcommand{\bx}{{\mathbf x}}
\newcommand{\cd}{{\cdot}}

\newcommand{\Omm}{\Omega_{{\rm m}}}

\begin{document}
\title{The low multipoles in the Pantheon+SH0ES data}

\author[]{Francesco Sorrenti,}
\author[]{Ruth Durrer and}
\author[]{Martin Kunz}
\affiliation[]{D\'epartement de Physique Th\'eorique and Center for Astroparticle Physics,\\
Universit\'e de Gen\`eve, 24 quai Ernest  Ansermet, 1211 Gen\`eve 4, Switzerland}
\date{today}

\emailAdd{francesco.sorrenti@unige.ch}
\emailAdd{ruth.durrer@unige.ch}
\emailAdd{martin.kunz@unige.ch}

\abstract{In previous work we have shown that the dipole in the low redshift supernovae of the Pantheon+SH0ES data does not agree with the one inferred from the velocity of the solar system as obtained from CMB data. We interpreted this as the presence of significant bulk velocities, indicating that it could be interesting to look at other large-scale multipoles. In this paper we study the monopole, dipole and quadrupole in the Pantheon+SH0ES data. We find that in addition to the dipole also both the monopole and the quadrupole are detected with high significance. They are of similar amplitudes as the bulk flow. While the monopole is only significant at very low redshift, the quadrupole even increases with redshift.}

\maketitle

\section{Introduction}
Standard cosmology assumes a statistically homogeneous and isotropic distribution of matter and radiation in the Universe. Correspondingly, on sufficiently large scales the geometry of the Universe is assumed to deviate little from homogeneity and isotropy, i.e., from a Friedmann-Lema\^\i tre (FL) universe.
These assumptions are in good agreement with the small fluctuations observed in the Cosmic Microwave Background (CMB), which is isotropic with  fluctuations of order $10^{-5}$, see~\cite{Planck:2018nkj,Planck:2018vyg,ACT:2020gnv,SPT-3G:2022hvq} for the latest results.

Due to our motion with respect to the surface of last scattering, the CMB also exhibits a dipole with an amplitude of about $10^{-3}$. This dipole has been discovered in the 1970s~\cite{1969Natur.222..971C,1971Natur.231..516H} and is now measured with exquisite precision~\cite{Kogut:1993ag,Planck:2013kqc,Planck:2018nkj}. This anisotropy in the CMB also leads to the correlation of adjacent multipoles which have consistently been measured with a significance of about 5 standard deviations~\cite{Saha:2021bay}. 
Attributing the entire CMB dipole to our motion, one infers a velocity of the solar system given by
\be
v_\odot=(369\pm 0.9){\rm km/s} \,, \qquad  
({\rm ra,dec}) =(167.942\pm 0.007, -6.944\pm 0.007)\,,
\label{e:d-Planck}
\ee
where $({\rm ra,dec})$ are the `right ascension' (ra) and `declination' (dec) denoting the directions with respect to the barycenter of the solar system (at J2000, i.e. January 1, 2000).
A possible intrinsic dipole in the CMB of the same order as the higher multipoles is expected to change this result by about 1\%. (This is simply a consequence of the fact that the clustering $C_\ell$'s are about four orders of magnitude smaller than $C_1$. Assuming that $C_1$ has a similar intrinsic clustering contribution this changes the dipole $\propto \sqrt{C_1}$ by about 1\%.)

Within the standard model of cosmology we expect to see this dipole due to our motion also in the large scale distribution of galaxies~\cite{Ellis:1984}. While first results of a radio survey agreed reasonably well with the CMB velocity~\cite{Blake:2002kx}, more recent analyses of catalogs of radio galaxies and quasars have found widely differing results from which significantly larger peculiar velocities have been inferred~\cite{Tiwari:2016,Bengaly:2017slg,Colin:2017juj,Secrest:2020has,Siewert:2021}. The latest results~\cite{Secrest:2022uvx} show a $5\si$ discrepancy with the CMB dipole which is considered by the authors as a challenge of the cosmological principle. There are, however also critiques that the analysis of the data might be too simplified and that a more refined analysis could give results that are consistent with standard cosmology~\cite{Dalang:2021ruy,Guandalin:2022tyl,Cheng:2023eba}, see also~\cite{daSilveiraFerreira:2024ddn} for an alternative method which gives results in agreement with the CMB dipole albeit with large error bars.

In previous work~\cite{Sorrenti_2023} we determined the dipole inferred from the Pantheon+ compilation of type Ia supernovae~\cite{Brout:2022vxf}. The first attempt to measure the dipole in supernova data dates back to 2006, however with a much smaller dataset and large error bars~\cite{Bonvin:2006en}. In~\cite{Sorrenti_2023} we found a dipole compatible in {\em amplitude} with the CMB dipole, but pointing in a different direction. However, the dipole amplitude in supernova distances is proportional to $[r(z)H(z)]^{-1}$ and hence it rapidly decays with redshift so that for supernovae with $z>0.1$ no significant dipole can be measured. Here $r(z)$ is the comoving distance out to redshift $z$ and $H(z)$ is the Hubble parameter at redshift $z$. It is therefore possible that the dipole we have seen in this data actually corresponds to the velocity $\bv_\odot-\bv^{\rm (bulk)}$ where $\bv_\odot$ is the peculiar velocity of the solar system and $\bv^{\rm (bulk)}$ is the bulk velocity of a sphere around us with radius $R\lesssim z/H_0=300(z/0.1)h^{-1}$Mpc, where $H_0$ is the present value of the Hubble parameter. Here we use the Hubble law for small redshifts, $z\ll1$ : the radius out to $z$ is given by $R(z)=z/H_0$ and $H_0=h/(3000$Mpc$)$. The speed of light is set to $c=1$ in our formulae. Interestingly, the bulk velocity inferred in this way is in relatively good agreement with the result of the CosmicFlows4 analysis~\cite{watkins23}, however our error-bars are significantly larger. In the present paper we test this hypothesis. As we discuss in the next section, if the SN dipole is really due to the peculiar motion of the individual supernovae and not due to a global dipole, we expect to observe also a monopole and a quadrupole (and higher multipoles) of similar amplitude. For this reason, we determine in this paper also the monopole and the quadrupole of the Pantheon+ compilation of supernova distances, in addition to the bulk flow (dipole) -- a possible quadrupolar Hubble expansion in Pantheon+ was also studied in~\cite{cowell_quadrupole} with the help of a cosmographic expansion. 

We do indeed find a monopole and a quadrupole with  amplitudes of the expected order of magnitude. We also argue that the amplitude we find for the bulk velocity is not extremely unlikely in the standard $\La$CDM model. 
\vspace{0.2cm}
\\
{\bf Notation :} We consider a spatially flat FL universe with linear scalar perturbations in Newtonian gauge,
\be
ds^2 = a^2(t)[-(1+2\Psi)dt^2 + (1-2\Phi)\de_{ij}dx^idx^j] \,.
\ee
The two metric perturbations $\Phi$ and $\Psi$ are the Bardeen potentials. Einstein's summation convention is assumed. Spatial vectors are denoted in bold face. The derivative with respect to conformal time  $t$ is indicated as an overdot.  $\HH=\dot a/a$ is the comoving Hubble parameter, while the physical Hubble parameter is given by $H=\dot a/a^2$. We work in units where the speed of light is unity but, for the convenience of the reader, we present our results on velocities in km/s.

\section{Theoretical description}\label{sec:theory}
In our previous paper~\cite{Sorrenti_2023}, we have found that even after subtracting the observer velocity $\bv_\odot$, assumed to be the one seen in the CMB data, there remains a significant dipole in the supernova distances of the Pantheon+ compilation.
We now want to study whether there are also significant monopole and quadrupole contributions. At first order in cosmological perturbation theory, the luminosity distance out to a source at observed redshift $z$ in direction $\bn$ is, up to some small local contributions which we neglect here, given by~\cite{Bonvin:2005ps,Hui_2006,Biern:2016kys}:
\bea
d_L(z,\bn) &=& \bar d_L(z)\Bigg\{1 - \frac{1}{\HH(z)r(z)}\bn\cd\bv_\odot -\Phi(\bn,z)
\nonumber \\ && \qquad
-\left(1-\frac{1}{\HH(z)r(z)}\right)\left[\Psi(\bn,z)+\bn\cd\bv(\bn,z)+\int_0^{r(z)}dr'(\dot\Psi+\dot\Phi)\right] \nonumber \\ &&  \qquad 
+\int_0^{r(z)}\frac{dr'}{r(z)}\left[1-\frac{r(z)-r'}{2r'}\De_\Om\right](\Phi+\Psi)\Bigg\} \label{e:21}
\,.
\eea
Here $\bv_\odot$ is the observer velocity and $\bv(\bn,z)$ is the peculiar velocity of the source. The functions are to be evaluated at $\bx=\bn r(z)$ and $t=t(z) = t_0-r(z)$. The symbol $\De_\Om$ denotes the Laplacian on the sphere, while $\bar d_L$ is the luminosity distance of the background FL universe. In \eqref{e:21} and in the following, the redshift $z$ is the observed, measured redshift. In the Pantheon+ data release this redshift is denoted $z{\rm HEL}$, indicating heliocentric redshift.

In a flat $\La$CDM universe at low redshift where radiation can be neglected it is given by
\be
\bar d_L(z)=(1+z)\int_0^z \frac{dz'}{H(z')}= \frac{1+z}{H_0}\int_0^z \frac{dz'}{\sqrt{\Omm(1+z')^3+1-\Omm}} \,.
\ee

\subsection{The monopole and quadrupole perturbations of the luminosity distance at low redshift}\label{sec:theory-quad}
In what follows we only retain the terms $\propto 1/[\HH(z)r(z)]$ in the perturbation of the luminosity distance. These terms dominate the fluctuations at small redshift. For $z\ll 1$ we may approximate $d_L$ by
\be\label{e:dL1}
d_L(z,\bn) = \bar d_L(z)\left[1 + \frac{1}{\HH(z)r(z)}\Big((\bv_\odot-\bv)\cd\bn\Big)\right] \,.
\ee
For small redshifts, $z\lsim 0.5$ this term dominates over the other contributions since it is enhanced by a factor $1/(r\HH)$ and since, at low redshift, velocities are about two orders of magnitude larger than the Bardeen potentials.
Using that  $ \bar d_L(z) =(1+z)r(z)$ et $\HH(z)=  H(z)/(1+z)$ we can write \eqref{e:dL1} as 
\be\label{e:dL2}
d_L(z,\bn) = \bar d_L(z) + \frac{(1+z)^2}{H(z)}\bn\cd\left(\bv_\odot-\bv(\bn,z)\right)\,.
\ee
If the source peculiar velocity $\bv(\bn,z)$ were independent of direction, a pure `bulk velocity' shared by all supernovae, this would lead to a pure dipole, which is what we considered in our previous paper. However, we expect that the peculiar velocity also depends on direction and on redshift. Here we assume that the redshift dependence is the one given by linear perturbation theory, this is a reasonable assumption for the large scales that we investigate (e.g. $z=0.025$ corresponds to a radius of $75h^{-1}$Mpc). We then fit for an angular dependence in the form of a monopole and a quadrupole. Of course we expect in principle also higher multipoles to be present but we neglect them here. As the different multipoles are orthogonal to each other, this should not bias our results on the monopole, dipole and quadrupole\footnote{{If the sky coverage would be homogeneous we would expect the different multipoles to be completely independent of each other. But since this is not the case, some bias might still be present.}}.
Within linear perturbation theory, the time dependence of the peculiar velocity field is given by
$$
\bv(\bx,z)=\frac{\dot D_1(z)}{\dot D_1(0)}\bv(\bx,0)\,,
$$
where $D_1$ is the linear growth function and the overdot denotes a derivative with respect to conformal time.
Introducing the growth rate $f(z)$ defined as~\cite{Durrer:2020fza}
\be
f(z) = -\frac{d\log(D_1)}{d\log(1+ z)} \,,
\ee
we can write
$$
\dot D_1(z) =\frac{D_1(z) f(z) H(z)}{(1+z)} \,.
$$
With this, \eqref{e:dL2} becomes
\be\label{e:dL3}
d_L(z,\bn) = \bar d_L(z) + \frac{(1+z)^2}{H(z)}\left[ \bn\cd\bv_{\rm \odot}-\frac{D_1(z) f(z)H(z)}{(1+z) D_1(0) f(0) H_0}\bn\cd\bv(\bn(t_0-t(z)),t_0) \right] \,.
\ee

As  mentioned above, if $\bv(\bn(t_0-t(z)),t_0)$ is independent of direction we obtain simply a dipole. Here we now go one step further by allowing for a dipole, a monopole and a quadrupole in the  directional dependence,
\be
\bn\cd\bv(\bn(t_0-t(z)),t_0) = \bn\cd\bv^{(\rm bulk)}  + n^i(\al_{ij}+\ga\,\de_{ij})n^j \,.
\ee
In this expression, $\al_{ij}$ is a symmetric traceless tensor which can be given, e.g. by the five components $\al_{11}$, $\al_{22}$, $\al_{12}$, $\al_{13}$ and $\al_{23}$. It represents the quadrupole of the peculiar velocity field today, and since $n^i\de_{ij}n^j=1$, $\ga$ corresponds to a monopole, the part of $\bv$ that is parallel to the radial direction $\bn$. Both, $\al_{ij}$ and $\ga$  have the units of a velocity. Putting all of this together we obtain
\be \label{theory:dl_quadrupole}
d_L(z,\bn) = \bar d_L(z) + \frac{(1+z)^2}{H(z)}\left[ \bn\cd\bv_{\rm \odot}-A(z)\left(\bn\cd\bv^{(\rm bulk)}  + n^i\,\al_{ij}\,n^j +\ga\right) \right]\, ,
\ee
where we defined the prefactor $A(z)$ as
\be \label{theory:prefactor}
A(z):=\frac{D_1(z) f(z)H(z)}{(1+z) D_1(0) f(0) H_0} \, .
\ee
The trace $\ga$ can be distinguished from $\bar d_L(z)$ via its redshift dependence $(1+z)^2A(z)/H(z)$ which differs from $\bar d_L(z)=(1+z)r(z)$. A short inspection shows that the ratio between these redshift dependencies, $(1+z)A(z)/(H(z)r(z))$, is well approximated by $1/z$ and  therefore becomes large at very low redshifts.

\begin{figure}[!ht]
\centering
  \includegraphics[scale=0.8]{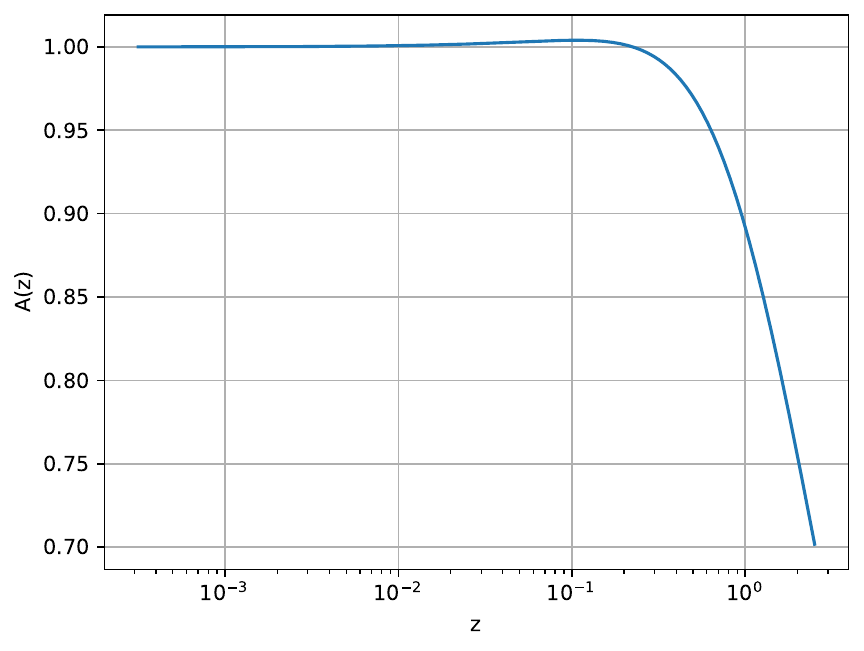}
    \caption{The prefactor $A(z)$ as function of  redshift. We assume $H_0=73.6$ km/s/Mpc and $\Omm=0.334$.}
  \label{fig:prefactor}
\end{figure}

In figure~\ref{fig:prefactor} we show the behaviour of the prefactor $A(z)$. At small redshifts, $z\lesssim 0.2$, we find that $A(z)\simeq 1$, implying that the correction due to this factor is negligible, in agreement with the assumption of a bulk motion of nearby galaxies.

In our assumption of a flat $\La$CDM universe, it is possible to obtain an analytical expression for the linear growth function and the growth rate:
\be \label{theory:fz1}
    D_1(z)=\frac{1}{5 \, (1+z) \, \Omm(0)}\left[ _2F_1 \left ( \frac{1}{3},1;\frac{11}{6};1-\frac{1}{\Om_\text{m}(z)} \right) \right],
\ee
where $_2F_1(a,b;c;d)$ represents the confluent hypergeometric function (see~\cite{Abra}, Chapter 13) and
\be \label{theory:fz}
    f(z)=\frac{1}{2}\Om_\text{m}(z) \left[ \frac{5}{_2F_1 \left ( \frac{1}{3},1;\frac{11}{6};1-\frac{1}{\Om_\text{m}(z)} \right) } - 3 \right],
\ee
with
\be
\Omm(z)=\frac{\Omm \, (1+z)^{3}}{\Omm \, (1+z)^{3}+(1-\Omm)}.
\ee

In our analysis we use the above expressions. In the literature, e.g.~\cite{Durrer:2020fza, Linder_2007}, it is common to find the approximate expression~\cite{Durrer:2020fza},
\be
f(z) \simeq \Omm^{0.56}\,,
\ee
which is in excellent agreement with \eqref{theory:fz}, as described in Appendix D of~\cite{Euclid:2023qyw}.

\subsection{Redshift corrections}\label{subsec:reds_corr}
The velocity-dependent terms in $d_L(z,\bn)$ actually stem from the fact that peculiar velocities modify the observed redshifts. They are the first terms in a Taylor series in $\delta z$. It might be more accurate to directly correct the redshift by subtracting $\delta z$ inside the expression for $\bar d_L(z)$. This is the method used in \cite{Carr_redshift_pantheon+} and we also adopt it here.  
The redshift correction due to the motion of the solar system is given by
\be \label{theory:z_cmb}
1+ z^{\text{(cmb)}}(z,\bn)=\frac{1+z}{1+z_\odot}\,,
\ee
where
\be \label{theory:z_sun}
1+ z_\odot=\sqrt{\frac{1 + (-v_\odot)/c}{1 - (-v_\odot)/c}} \,,
\ee
with $v_\odot=\bn\cd\bv_\odot$.

Similarly, we consider the redshift correction due to the peculiar motion of the supernovae relative to the solar system which we model as a bulk velocity, a monopole and a quadrupole,
\be \label{theory:z_quadrupole}
1+z_{q}(z,\bn)=\frac{1+z^{\text{(cmb)}}}{1+z_{p}(z,\bn)} =\frac{1+z}{(1+z_\odot)(1+z_{p}(z,\bn))}\,,
\ee
where, similar to \eqref{theory:z_sun}, $z_p$ is given by
\be \label{theory:z_p}
1+ z_{p}=\sqrt{\frac{1 + (v_{p})/c}{1 - (v_{p})/c}} \,,
\ee
with $v_p$ given by
\be \label{theory:v_p}
v_p = A(z)\left[\bn\cd\bv^{(\rm bulk)}  + n^i\al_{ij}n^j +\ga\right] \, .
\ee
For \eqref{theory:z_quadrupole} we simply use the formula for the redshift combining two boosts, one to the CMB frame and one due to the peculiar velocity of the source. We can then rewrite \eqref{theory:dl_quadrupole} more concisely as
\be \label{theory:dl_z_quadrupole}
d_L(z,\bn) = \bar d_L(z_{q}(z,\bn))  \,.
\ee

\section{Data and methodology}

As in our previous work~\cite{Sorrenti_2023}, we use the Pantheon+ data which provides distance moduli $\mu$ for 1550 SNe,
\be
\mu = 5\log_{10}(d_L/10{\rm pc}) =  5\log_{10}(d_L/1{\rm Mpc}) +25 \,.
\label{e:distance_modulus}
\ee
As reported in table~\ref{t1}, 77 SNe are in galaxies that also host Cepheids, for which we know the absolute distance modulus $\mu_\mathrm{ceph}$. While Pantheon+ uses corrected redshifts including the motion of the solar system and estimated peculiar velocities of the sources, we use the actually measured redshifts for our analysis.

For handling the astronomical quantities and convenient unit conversions we use the package \texttt{astropy}~\cite{astropy:2013,astropy:2018,astropy:2022}. The theoretical model introduced in section~\ref{sec:theory} is implemented in the code \texttt{scoutpip}\footnote{Available at \url{https://github.com/fsorrenti/scoutpip}}.
To determine the parameters of our model we perform an MCMC analysis using the python package \texttt{emcee}~\cite{emcee}. Our code is parallelized using the Python package \texttt{schwimmbad}~\citep{schwimmbad}. Our sampler consists of 32 walkers with the ``stretch move" ensemble method described in~\cite{autocorr}. 

As in our previous paper, we maximize the likelihood
\be \label{eq:likelihood}
\log(\mathcal{L}) = - \frac{1}{2} \Delta \boldsymbol{\mu}^T C^{-1} \Delta \boldsymbol{\mu},
\ee
where $C$ is the covariance matrix provided by the Pantheon+ collaboration\footnote{We use the full covariance matrix, the sum of statistical and systematic errors.}. The vector $\Delta \boldsymbol{\mu}$ is defined by
\be \label{e:cases}
\Delta \mu^i =
\begin{cases}
\mu^i + dM -\mu_\mathrm{ceph}^i,  \quad i \in  \text{Cepheid hosts}\\
\mu^i +  dM -\mu_\mathrm{model}^i, \quad \text{otherwise} \\
\end{cases}
\ee 
where
\be
\mu^i_\mathrm{model} = 5 \log \biggl( \frac{d_L(z_i, \bn_i)}{\rm Mpc} \biggr) + 25 \, ,
\ee
and $d_L(z_i, \bn_i)$ is given in \eqref{theory:dl_z_quadrupole}.
In eq.~\eqref{e:cases} we introduce the nuisance parameter dM which is constrained by the supernovae in Cepheid-host galaxies, while the supernovae in galaxies not hosting Cepheids constrain the luminosity distance parameters.

Finally, we analyse our chains using the \texttt{getdist} package~\cite{getdist}.
Following the \texttt{emcee} guidelines (the interested reader is referred to \url{https://emcee.readthedocs.io/en/stable/tutorials/autocorr/}), we use the integrated $\tau$~\cite{autocorr} as convergence diagnostics.
Here $\tau$ can be considered as the number of steps necessary for the chain to forget where it started. In particular, we assume that the chain is converged with respect to a certain parameter when the number $N$  of steps in the chain is larger than 50 times the  auto-cor\-re\-lation time, $N > 50 \tau$.
We further consider as burn-in and discard the first  $2 \, \lfloor \tau_{\rm max} \rfloor$ steps, where $\tau_{\rm max}$ is the maximum $\tau$ value for all the parameters.
In all MCMC analyses, we use  uniform priors as reported in table~\ref{t:prior} for the parameters that are varied.
\begin{table}[!ht]
\centering
\begin{tabular}{ c c }
\toprule
 Parameter & Prior range  \\
 \midrule
 $|\bv^{(\rm bulk)}|$ & [0, 1000] km/s \vspace{2pt}\\
 \text{ra} & [0\textdegree, 360\textdegree] \vspace{2pt}\\
 $\sin(\text{dec})$ & [-1, 1]  \vspace{2pt}\\
 $\al_{ij}$ & [-500,500]  km/s \vspace{2pt}\\
 $\ga$  & [-500,500]  km/s \vspace{2pt}\\
 dM & [-1, 1] \vspace{2pt}\\  
 $H_0$ & [40, 100] km/s/Mpc  \vspace{2pt}\\  
 $\Omm$ & [0, 1]  \\
 \bottomrule
 \end{tabular}
 \caption{Uniform priors for the parameters sampled in the various MCMC analyses. We vary $\sin(\text{dec})$ and not dec in order to  sample the celestial sphere uniformly in area. We then apply the $\arcsin$ function to the chain entries in order to recover the declination for the analysis. $\al_{ij}$ refers to a generic element of the quadrupole matrix introduced in eq.~\eqref{theory:dl_quadrupole}.
 \label{t:prior}}
 \end{table}

\newpage
\section{Results}

\subsection{Simple dipole analysis} \label{subsec:simple_dipole} 
As first step, in order to test our code and the theoretical assumptions, we perform a similar analysis for the dipole only as in our previous paper~\cite{Sorrenti_2023}, but applying the redshift correction as described in section~\ref{subsec:reds_corr}, neglecting peculiar velocity corrections. 
\begin{figure}[!ht]
\centering
  \includegraphics[scale=0.5]{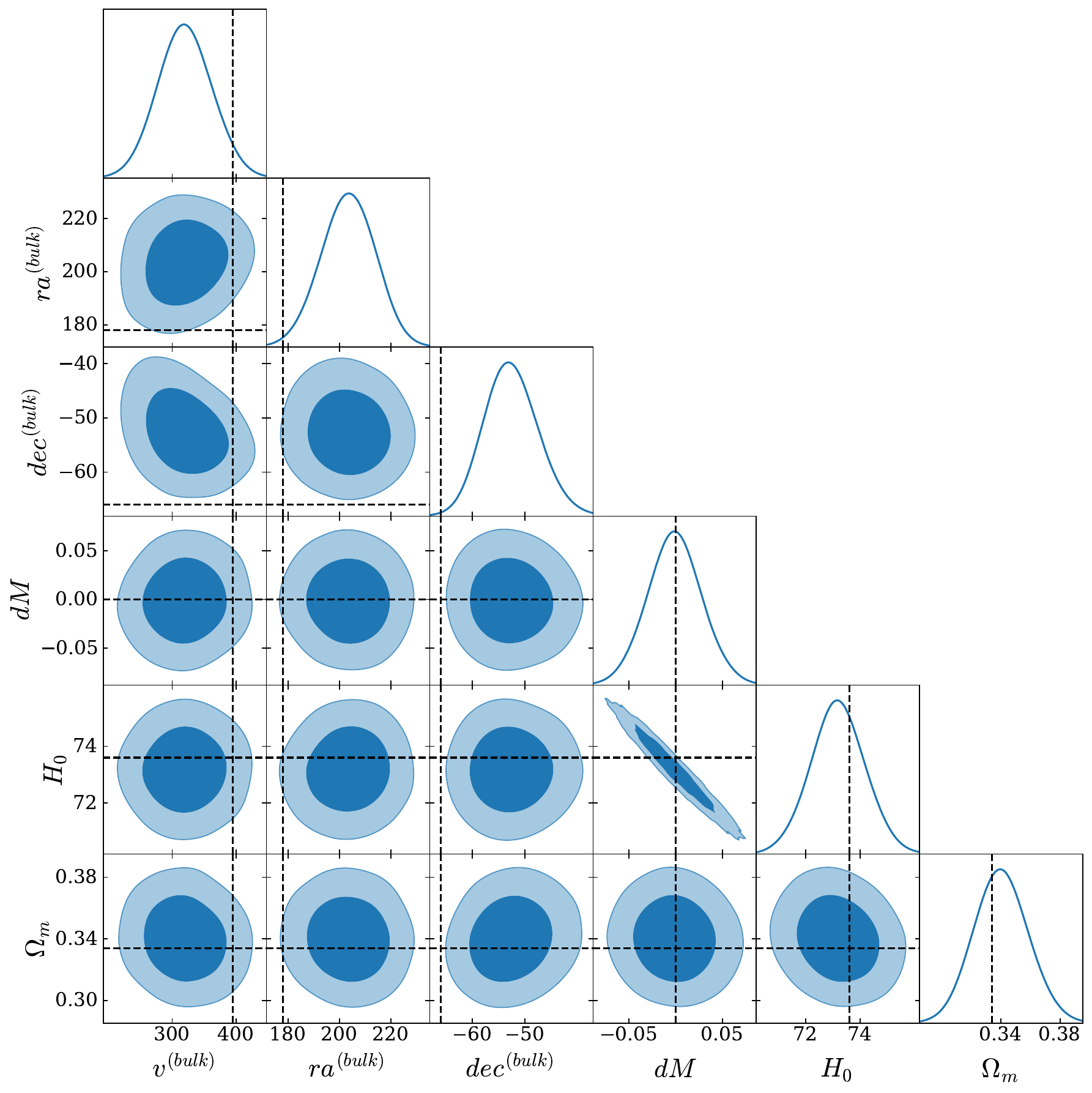}
    \caption{Contour plots for the redshift correction as described in section~\ref{subsec:reds_corr}, including just the $\bv^{\text{(bulk)}}$ correction in equation~\eqref{theory:z_p}, i.e. considering only the first term in brackets in equation~\eqref{theory:v_p} and fixing $\bv_\odot$ to \eqref{e:d-Planck} according to Planck. The dashed lines show as reference value for $\bv^{\text{(bulk)}}$ the bulk flow obtained in the CosmicFlows4 project, see  \cite{watkins23}, for a sphere of radius R=150$h^{-1}$Mpc (i.e. $| \bv^{\text{(bulk)}}|=395$km/s, ra$^{(\rm bulk)}=178$\textdegree, dec$^{(\rm bulk)}=-66$\textdegree), dM$=0$, $\Omm=0.334$ and $H_0=73.6 ($km/s/Mpc$)$ as obtained by Pantheon+~\cite{Brout:2022vxf}.}
  \label{fig:only_bulk}
\end{figure}

Doing this we obtain the same results as in our previous paper where the corrections were applied at the level of the definition of luminosity distance instead of the redshift, i.e., using only the first term in the Taylor series in $\bar d_L(\bar z+\de z)$ and linearising $\de z$ in $\bn\cd\bv_\odot$ (neglecting bulk velocity, the monopole and the quadrupole). 
The robustness of applying redshift corrections is also manifest by the fact that, when considering the dipole only, we find a negligible difference of $\Delta \chi^2 \simeq 0.8$ with respect to the analysis developed in the previous paper where we used eq.~\eqref{theory:fz}.
In the first row of table~\ref{tab:common_table} we report the constraints inferred from the MCMC routine for the new orange contours, where the bulk velocity correction is taken into account as a redshift correction. Note that while the amplitude roughly agrees with the velocity of the solar system inferred from the CMB, the direction is very different, compared with eq.~\eqref{e:d-Planck}. This result is in excellent agreement with our previous paper~\cite{Sorrenti_2023}, where it was the main finding. It led us to the conclusion that the bulk velocity cannot be neglected.

\subsection{Including a bulk velocity} 
\label{sec:bulk_only_analysis}
When fixing $\bv_\odot$ to the Planck value and including just the $\bv^{(\rm bulk)}$ correction in equation~\eqref{theory:z_p}, i.e.\ considering only the first term in brackets in equation~\eqref{theory:v_p}, the dipole, we obtain the contour plots shown in figure~\ref{fig:only_bulk} and reported in the second row of table~\ref{tab:common_table}. These bulk velocities agree with our previous paper~\cite{Sorrenti_2023}. As discussed there in details, the direction of the fitted bulk flow agrees well with the bulk flow direction assumed in the Pantheon+ analysis~\cite{Brout:2022vxf}, but the amplitude is nearly twice as large.

\subsection{Bulk + quadrupole analysis}

\begin{figure}[h!]
    \centering
  \includegraphics[width=\linewidth]{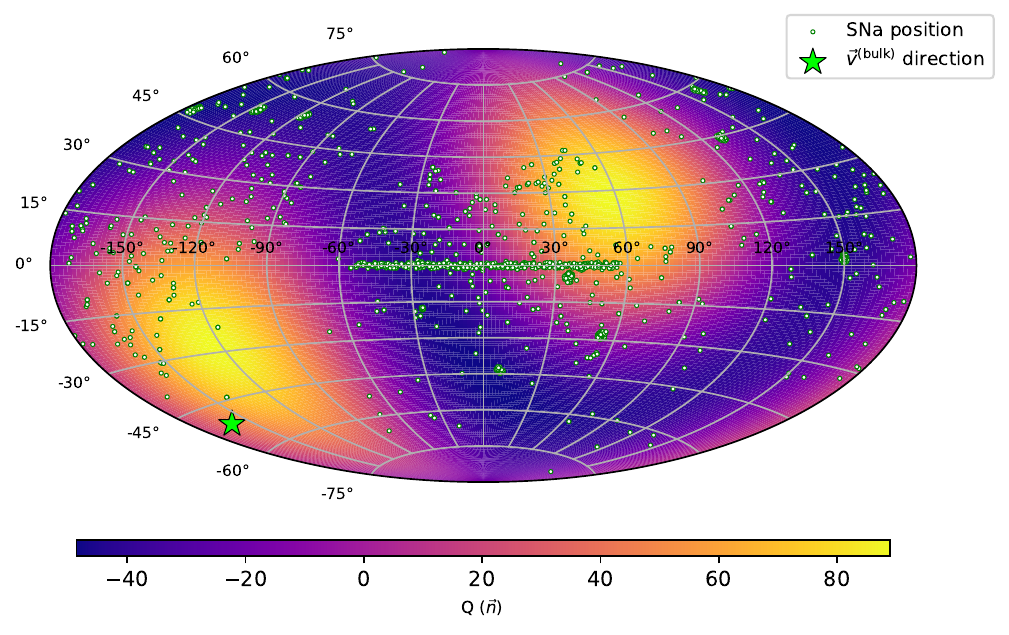}
    \caption{We show  the inferred quadrupole function, $Q(\bn)=\al_{ij}n^in^j$ in equatorial coordinates. The Supernova positions as well as the direction of the bulk velocity are also indicated. Note that the peak of the quadrupole is not far from the bulk velocity direction. No strong correlation with the supernova positions is evident.}
\label{fig:quadrupole_function_position}
\end{figure}

We now include also the quadrupole in the luminosity distance, which  comes from the angular dependence of the peculiar velocity field as discussed in section~\ref{sec:theory-quad} and which we describe by the matrix $(\al_{ij})$, to perform a \textit{bulk + quadrupole} analysis. As mentioned in section~\ref{sec:theory-quad}, $(\al_{ij})$ is a symmetric trace-less tensor of dimension $3 \times 3$. It is defined by  five parameters (e.g. $\al_{11},~\al_{22},~\al_{12},~\al_{13}$ and $\al_{23}$) that we introduce in our MCMC routine.
In Fig.~\ref{fig:quadrupole_function_position} we visualise the quadrupole contribution plotting the function $Q(\bn)=\al_{ij}n^in^j$.

Instead of the matrix elements $\al_{ij}$, however, we show in our plots the geometrically more interesting quantities given by the eigenvalues and the direction of the eigenvectors. Note that these also amount to five parameters given e.g. by $\lambda_1$ and $\lambda_2$, the traceless condition then determines $\lambda_3$, as well as three angles determining the direction of the first two orthogonal eigenvectors. The third eigenvector is then simply given by the orthogonality condition, $\bw_3=\bw_1\times\bw_2$. Note that with $\bw_i$ also $-\bw_i$ is an eigenvector, so we can fix their orientation at will.

\begin{minipage}{0.45\linewidth}
\begin{figure}[H]
\centering
  \includegraphics[width=0.9\linewidth]{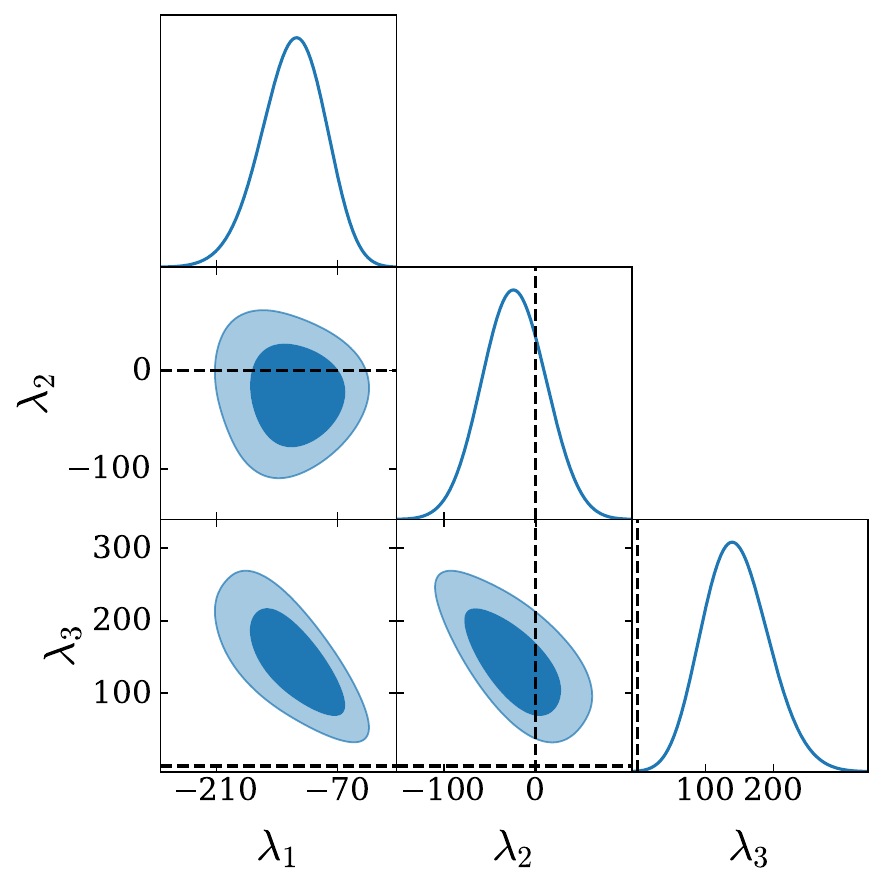}
    \caption{Distribution of the eigenvalues of $(\al_{ij})$ without redshift cut. Dashed lines indicate the values $\lambda_i=0$.}
\label{fig:eigenvalues_quadrupole_no_filter}
\end{figure}
\end{minipage}~~
\begin{minipage}{0.45\linewidth}
~~ 
\begin{figure}[H]
    \centering
  \includegraphics[width=0.9\linewidth]{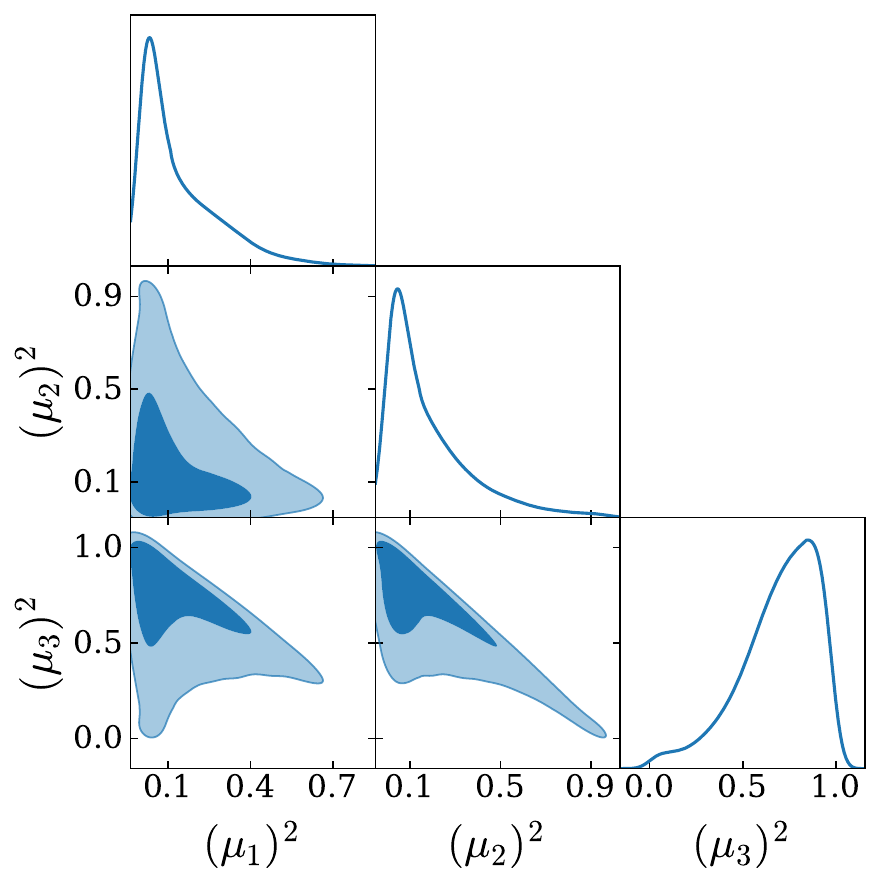}
    \caption{The scalar product $\mu_i^2=\left[ (\bw_i \cdot \bv^{\text{(bulk)}})/| \bv^{\text{(bulk)}}| \right]^2$, analysis without redshift cut.}
\label{fig:bulk_scalar_eigenvector}
\end{figure}
\end{minipage}
\vspace{12pt}

Using the samples obtained for the matrix elements $\al_{ij}$, we determine the contour plots for the eigenvalues which are shown in figure~\ref{fig:eigenvalues_quadrupole_no_filter} and reported in table~\ref{tab:eigenvalues}. 
\begin{table}[!ht]
    \centering
\begin{tabular}{c c c}
{$\lambda_1$ \small{[km/s]}     } & {$\lambda_2$ \small{[km/s]}    } & {$\lambda_3$  \small{[km/s]}  } \vspace{2pt}\\
\midrule \vspace{2pt}
$-121^{+39}_{-33}$& $-23\pm 34$& $145^{+40}_{-50}$ \\
\end{tabular}
\caption{Posteriors for eigenvalues of the quadrupole matrix $(\al_{ij})$.}\label{tab:eigenvalues} 
\end{table}

In our parametrisation of the matrix $(\al_{ij})$ the trace vanishes, hence the sum of the eigenvalues is zero by construction. However, eigenvalues $\lambda_1$ and $\lambda_3$ differ from zero by 3 to 4 standard deviations. Since their distributions are close to Gaussian, see figure~\ref{fig:eigenvalues_quadrupole_no_filter},  we conclude that the detection of the quadrupole is significant. The direction of the eigenvectors is fixed up to a sign. To remove this ambiguity, we choose all eigenvectors to point into the  
 northern hemisphere. They are normalized and dimensionless, since we assume the eigenvalues to have the dimension of velocity. In table~\ref{tab:eigenvectors} we report their directions. As $\lambda_2$ is compatible with zero, it is not surprising that the direction of $\bw_2$ is not well determined. However, also the direction of $\bw_1$ has surprisingly large errors. 
 
 To compare the amplitude of the quadrupole with the bulk flow, we define
 \be
 \lambda=\sqrt{\lambda_1^2+\lambda_2^2+\lambda_3^2} \,.
 \ee
 and compare it with $|\bv^{(\rm bulk)}|$ obtained in the \textit{bulk + quadrupole} analysis. While $|\bv^{(\rm bulk)}| = 338 \pm 40 \, $km/s, we find  $\lambda = 190 \pm 40 \, $km/s.\footnote{Errors associated to $\lambda$ are computed using the python package \texttt{uncertainties}~\cite{uncertainties}, which handles error propagation to determine the errors of functions of quantities with uncertainties.} Even though the quadrupole is somewhat smaller than the dipole, it is of a comparable order of magnitude.


\begin{table}[!ht]
    \centering
\begin{tabular}{c c c c}
{} & {ra [\textdegree]} & {dec [\textdegree]} & $(\mu_i)^2$~[(km/s)$^2$]  \vspace{2pt} \\
\midrule
$\bw_1$& $230\pm 60$& $38\pm{20}$ & $0.132^{+0.068}_{-0.17}$\vspace{6pt}\\
$\bw_2$& $242^{+90}_{-80}$& $25^{+10}_{-20}$ & $0.171^{+0.083}_{-0.21}$\vspace{6pt}\\
$\bw_3$& $71^{+10}_{-30}$& $37\pm 10$ & $0.70^{+0.25}_{-0.14}$\\
\bottomrule
\end{tabular}

\caption{\label{tab:eigenvectors} 
Position of the eigenvectors $\bw_i$ in the northern hemisphere. In the last column we also show the scalar product $(\mu_i)^2=\left[ (\bw_i \cdot \bv^{\text{(bulk)}})/| \bv^{\text{(bulk)}}| \right]^2 $.}
\end{table}

In figure~\ref{fig:bulk_scalar_eigenvector} and table~\ref{tab:eigenvectors} we also show the scalar products $\mu_i^2=\left[ (\bw_i \cdot \bv^{\text{(bulk)}})/| \bv^{\text{(bulk)}}| \right]^2 $. Note that the sign has no significance since both $\bw_i$ and $-\bw_i$ are eigenvectors of $\lambda_i$. Since the direction of $\bw_2$ is not well determined the value of $\mu_2$ is also not. Interestingly, $\bw_3$ is well aligned with $\bv^{\rm(bulk)}$ and, as a consequence,  $\bw_1$ is nearly orthogonal to $\bv^{\rm(bulk)}$.

\begin{figure}
    \centering
  \includegraphics[width=0.9\linewidth]{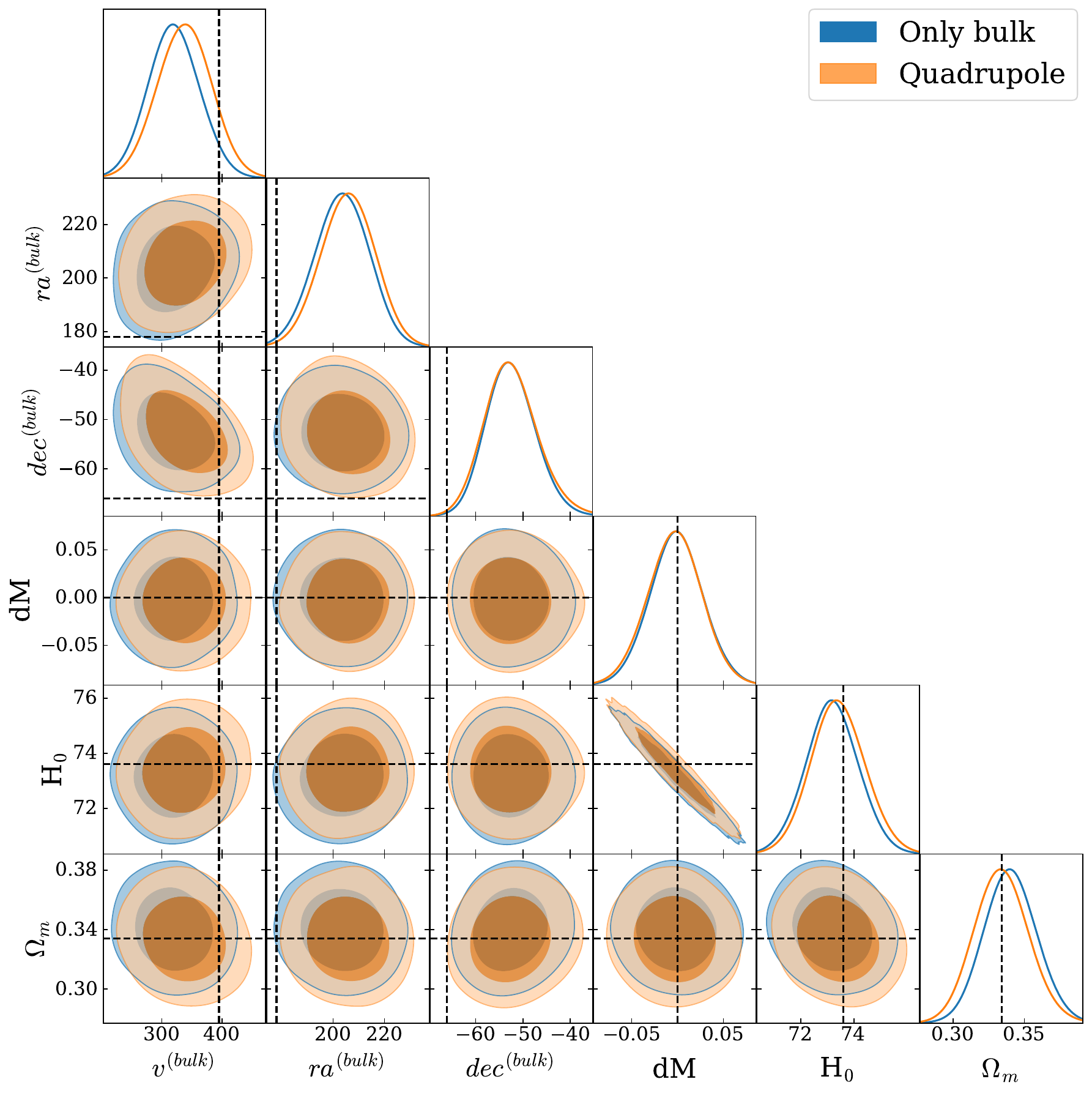}
    \caption{ In this figure we show the parameters not involving the quadrupole and the monopole and compare their values in an analysis including the quadrupole (orange) and an analysis containing only the dipole from the bulk velocity.}
\label{fig:only_bulk_vs_quadrupole}
\end{figure}

In figure~\ref{fig:only_bulk_vs_quadrupole} we compare the inferred cosmological parameters and the bulk velocity in an analysis including the quadrupole from the bulk motion (orange) with the one including only the dipole. We find that all values are virtually identical in both analyses. It is not surprising that the dipole is not changed, as we expect different multipoles to be independent, but we note that also the cosmological parameter constraints stay the same.

\subsection{Bulk + monopole analysis\label{subsection:bulk_monopole_analysis}}
We also model the data by adding a monopole to the bulk velocity, the parameter $\gamma$ of eq.~\eqref{theory:v_p}. 
The unperturbed cosmological $\bar d_L(z)$ of course also represents a monopole. But the redshift dependence of the velocity monopole, given by $(1+z)^2H^{-1}(z)A(z)$  is quite different from $\bar d_L(z)$ so that this degeneracy is lifted.  Note especially that contrary to $\bar d_L(z)$, the redshift contribution from the velocity monopole does not vanish for $z\ra 0$.  This is due to the fact that a radial velocity of the source modifies the measured redshift so that $d_L=0$ now no longer coincides with vanishing redshift. This makes the monopole at low redshift very distinct from $\bar d_L(z)$.

Interestingly, while adding a quadrupole with its five free parameters reduces the $\chi^2$ by the modest value $\Delta \chi_Q^2 \simeq 5.9$ with respect to the analysis including only a bulk velocity of section~\ref{sec:bulk_only_analysis}, by adding a monopole characterized by just one free parameter, $\gamma$, we gain a $\Delta \chi^2_M \simeq 6.85$. The inclusion of the monopole also leads to a slight increase of $H_0$, by $0.9\sigma$ and to a decrease of $\Omm$ by about $1.2\si$, see third row of table~\ref{tab:common_table}. 
The increase of $H_0$ can be understood as follows: Considering eqs. \eqref{theory:z_quadrupole} to\eqref{theory:v_p} we see that a negative value of $\ga$ leads to an increase in $z_q$ with respect to its value for $\ga=0$ hence
the measured $d_L(z) =\bar d_L(z_q)> \bar d_L(z)$. And since $H_0$ is inversely proportional to $\bar d_L(z)$, this implies a larger $H_0$. At $z<0.01$ this reduction of $\bar d_L(z)$ is about 1.2\%, but due to the reduced value of $\Omm$, it decays rapidly and is only about 0.02\% at $z=0.5$.   The difference $\bar d_L(z,\Omm=0.316,$ $H_0=74.2)-\bar d_L(z,\Omm=0.34,$ $H_0=73.2)$ crosses zero at $z\simeq 1$, above which the Pantheon+ dataset contains no SNIa.

However, like the quadrupole the monopole does not affect the dipole, $\bv^{(\rm bulk)}$. This again is a consequence of the fact that the different multipoles are orthogonal functions. As the background luminosity distance is a monopole, only the monopole of the peculiar velocity, i.e. its radial component, can affect it and thereby modify the inferred cosmological parameters.
 
\begin{figure}
\centering
\includegraphics [scale=0.4]{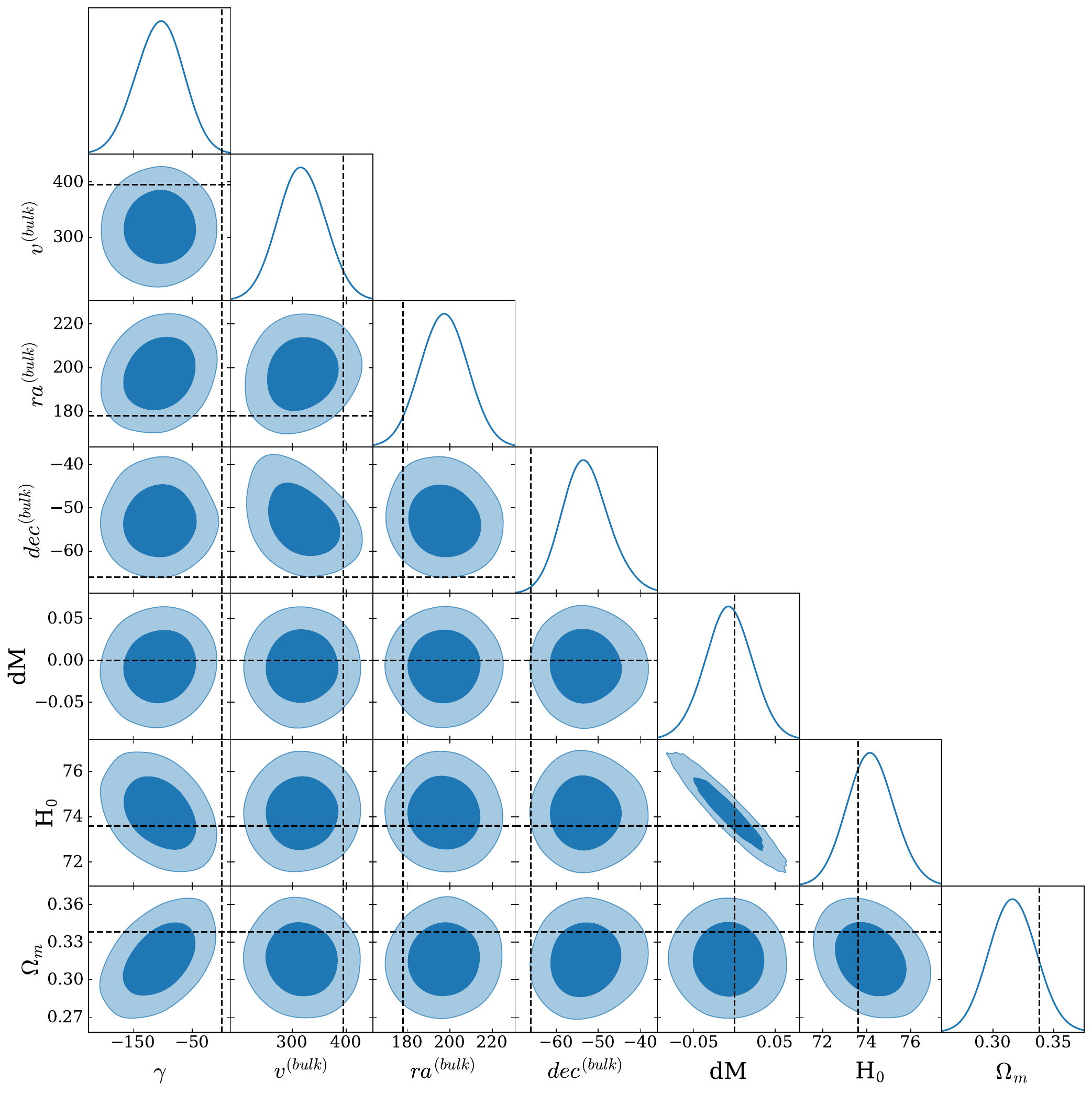}
	\caption{Contour plots for the  analysis including a bulk velocity and monopole, corresponding to a radial peculiar velocity. \label{fig:contours_monopole_bulk}
}
\end{figure}
\begin{table}[!ht]
    \setlength\tabcolsep{4 pt} 
    \scriptsize
\centering
    \begin{tabular}{cccccccccc}
        \toprule \vspace{2pt}
        Model & $\gamma$& $|\bv_\odot|$& ra$_\odot$& dec$_\odot$ & $|\bv^{(\rm bulk)}|$& ra$^{(\rm bulk)}$& dec$^{(\rm bulk)}$ &  $H_0$ & $\Omm$ \\         
        &\scriptsize{[km/s]}&\scriptsize{[km/s]}&\scriptsize{[\textdegree]}&\scriptsize{[\textdegree]}&\scriptsize{[km/s]}&\scriptsize{[\textdegree]}&\scriptsize{[\textdegree]}&\scriptsize{[km/s/Mpc]}& \\
		\midrule 
Sec.~\ref{subsec:simple_dipole} & - & $318\!\pm\! 40 $ & $140\!\pm\! 7.7$ & $42^{+7}_{-6}$& - & - & -  &$73.2\pm 1.0   $ & $0.340\pm 0.018$  \vspace{6 pt} \\
Sec.~\ref{sec:bulk_only_analysis} & - & - & - & - & $320\pm 40 $ & $203\pm 11$ & $-52.5^{+4.8}_{-5.5}$ &  $73.2\pm 1.0   $ & $0.340\pm 0.018$  \vspace{6 pt} \\
Sec.~\ref{subsection:bulk_monopole_analysis} & $-106\!\pm\! 40 $ & - & - & - & $318\pm 40 $ & $197\pm 11  $ & $-52.8^{+5.0}_{-6.0}$ & $74.2\pm 1.1$ & $0.316\pm 0.019$ \vspace{6 pt} \\
    \bottomrule
    \end{tabular}
 \\
     \caption{Constraints on parameters for the models we discussed in sections~\ref{subsec:simple_dipole},~\ref{sec:bulk_only_analysis} and~\ref{subsection:bulk_monopole_analysis} without imposing any redshift cut.  For the sake of simplicity in presenting the results, we omit the constraints on dM. Here and in all the following results tables, the errors show the 68\% confidence intervals obtained when analysing the MCMC chains with \texttt{getdist}. They are purely statistical errors, as a consequence, they should be interpreted with care. \label{tab:common_table}}
\end{table}

The contour plots for this analysis are shown in figure~\ref{fig:contours_monopole_bulk}. Note the correlation between $\ga$ and $\Omm$ and the anti-correlation of $\ga$ and $H_0$. Despite the relatively low mean value amplitude of the monopole, about $1/3$ of the bulk velocity, its impact on the cosmological parameters is quite strong.
Despite the fact that the monopole is distinguished from the background luminosity distance only via its redshift dependence, it is detected with a significance of more than $2\si$. As we shall see, this is mainly due to its strong effect at very low redshift.

\subsection{The full bulk + quadrupole + monopole analysis\label{subsection:full_analysis}}
Finally we model the redshift by adding all, the radial velocity (monopole), the bulk velocity (dipole) and the quadrupole. With respect to the analysis allowing only for a bulk velocity we performed in section~\ref{sec:bulk_only_analysis}, we gain a $\Delta \chi^2 \simeq 9.31$ in this model which has six additional parameters. As we have seen in the previous section, most of this improvement is due to the monopole. The $\De\chi^2$ for the quadrupole alone is only somewhat larger than 5 which is the expected value for 5 new parameters and not a truly better fit, while for the single parameter $\ga$,  $\chi^2$ is reduced by more than 6.

\begin{figure}
\includegraphics [scale=0.3]{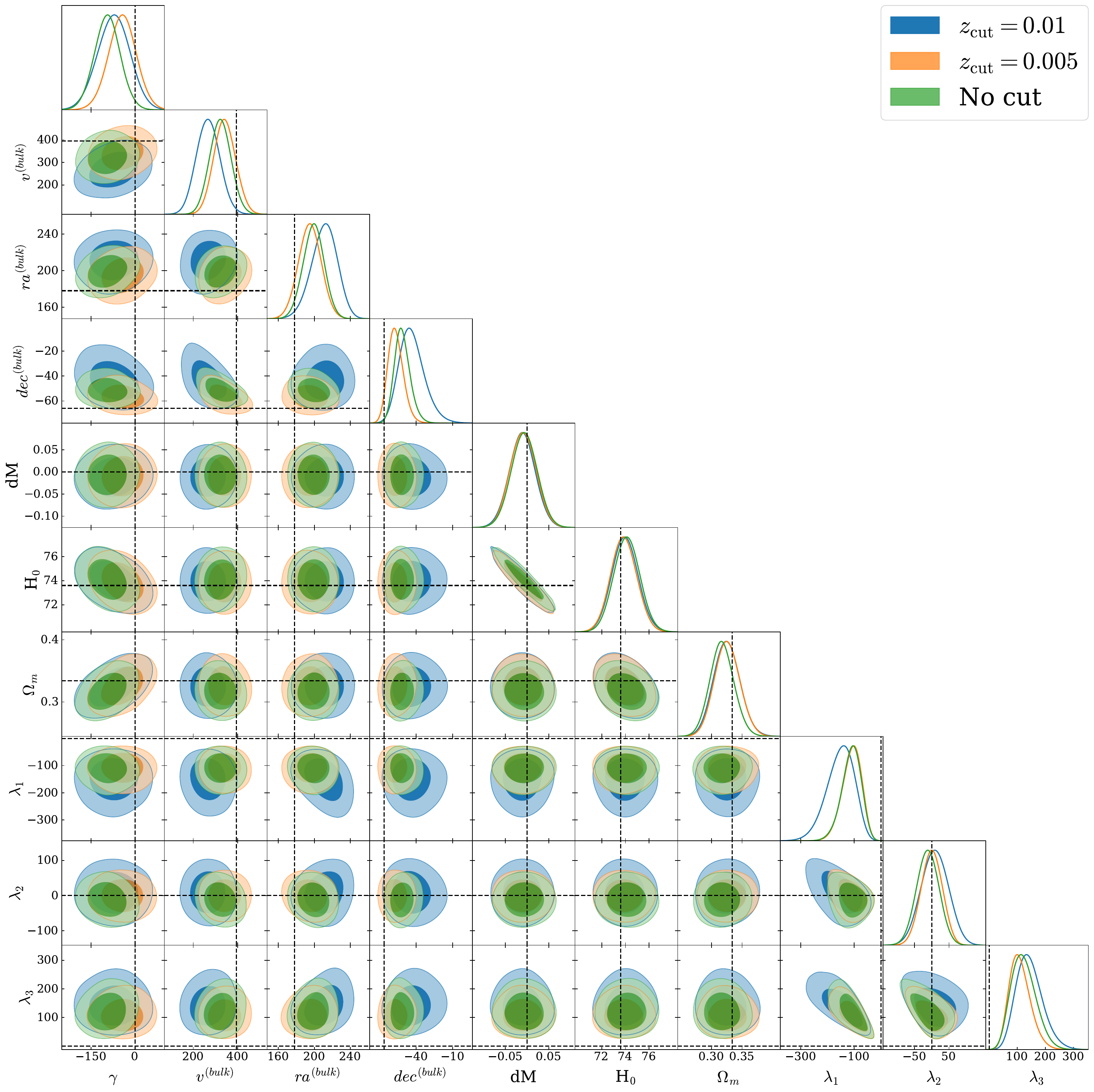}
	\caption{Contour plots for the full bulk velocity, quadrupole and monopole analysis described in section~\ref{subsection:full_analysis} with three different cuts in the redshift of the supernovae. More cuts are shown in figure~\ref{fig:med_full_analysis} in Appendix~\ref{appendix:extra_plots}.} \label{fig:contours_monopole_bulk_quadrupole_no_filter}
\end{figure}

The contour plots of this analysis are shown in 
figure~\ref{fig:contours_monopole_bulk_quadrupole_no_filter} (green contours and lines) and the mean values with $1\si$ error bars are reported in table~\ref{tab:params_monopole_bulk_quadrupole} (first line).  As already in the pure monopole analysis, the monopole amplitude, $\ga$ is correlated with $\Omm$ and $H_0$. The mean values of $H_0$ and $\Omm$ are affected by the presence of $\ga$, but the values found in the original Pantheon+ analysis~\cite{Brout:2022vxf} remain consistent within $1\si$ with our results.

\begin{figure}
\centering
\includegraphics [scale=0.55]{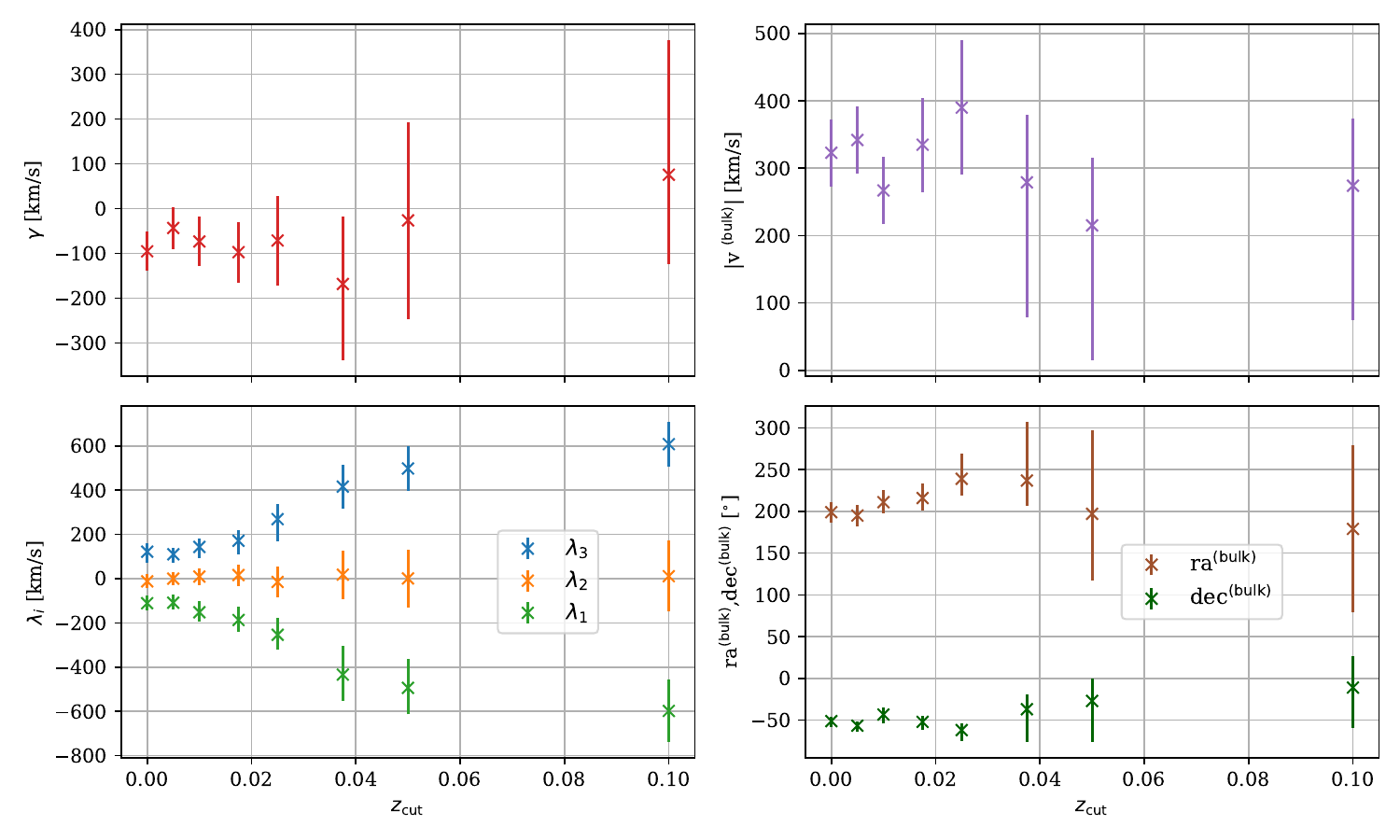}
	\caption{Visualisation of the constraints on monopole, dipole and quadrupole as reported in table~\ref{tab:params_monopole_bulk_quadrupole}.} \label{fig:summary_constraints}
\end{figure}
\begin{table}[!ht]
    \setlength\tabcolsep{3.5 pt} 
    \scriptsize
    \begin{tabular}{cccccccccc}
        \toprule
		 z$_{\rm cut}$ &$\gamma$ &$|\bv^{(\rm bulk)}|$& ra$^{(\rm bulk)}$& dec$^{(\rm bulk)}$ & $H_0$& $\Omm$& $\lambda_1$ & $\lambda_2$ & $\lambda_3$ \\ 
		&\scriptsize{[km/s]} &\scriptsize{[km/s]}&\scriptsize{[\textdegree]}&\scriptsize{[\textdegree]}&\scriptsize{[km/s/Mpc]}& &\scriptsize{[km/s]}&\scriptsize{[km/s]}&\scriptsize{[km/s]} \\
  \midrule
		No cut & $-95\pm 44 $ & $323\pm 50 $ & $199\pm 12  $ & $-51.2^{+5.4}_{-7.0}$ & $74.1\pm 1.1$ & $0.317\pm 0.019$ & $-111^{+40}_{-31}$ &$-12\pm 32$ & $122^{+40}_{-50}$\vspace{6 pt} \\
    	0.005 & $-43\pm 47$ & $342\pm 50$ & $195\pm 13$ & $-56.6^{+5.4}_{-6.8}$ & $73.9\pm 1.1$ & $0.326\pm 0.021 $ & $-109^{+40}_{-31}  $ &$0\pm 30   $ & $110^{+30}_{-40}$\vspace{6 pt} \\
  		0.01 & $-73\pm 55 $ & $267\pm 50 $ & $211^{+15}_{-13} $ & $-43.3^{+8.5}_{-11}  $ & $74.0\pm 1.1$ & $0.325\pm 0.021$ & $-152^{+58}_{-43} $ &$9\pm 39$ & $143^{+40}_{-50}$\vspace{6 pt} \\
  		0.0175 & $-97\pm 68 $ & $335\pm 70$ & $216^{+17}_{-15}$ & $-52.1^{+7.2}_{-9.9}  $ & $74.1\pm 1.1 $ & $0.323\pm 0.023$ & $-187^{+69}_{-53}$ &$16\pm 48$ & $171^{+50}_{-60}$\vspace{6 pt} \\
  		0.025 & $-71\pm 100 $ & $390\pm 100$ & $239^{+30}_{-20}$ & $-61.9^{+8.0}_{-13}$ & $74.0\pm 1.1$ & $0.323\pm 0.024$ & $-254^{+87}_{-68}$ &$-15\pm 69$ & $269^{+70}_{-100}$\vspace{6 pt} \\
  		0.0375 & $-168^{+150}_{-170}$ & $279^{+100}_{-200}$ & $237^{+70}_{-30} $ & $-37^{+18}_{-39}  $ & $74.0\pm 1.2 $ & $0.322^{+0.025}_{-0.028} $ & $-434^{+140}_{-120} $ &$18\pm 110 $ & $416\pm 100$\vspace{6 pt} \\
  		0.05 & $-26\pm 220  $ & $215^{+100}_{-200}$ & $197^{+100}_{-80}$ & $-27^{+27}_{-49}$ & $73.8\pm 1.3  $ & $0.327^{+0.027}_{-0.031}$ & $-498^{+140}_{-120}$ &$1\pm 130$ & $498\pm 100$\vspace{6 pt} \\
  		0.1 & $76^{+300}_{-200}$ & $274^{+100}_{-200}$ & $179\pm 100$ & $-11^{+38}_{-48}$ & $73.4\pm 1.3  $ & $0.337^{+0.027}_{-0.030}$ & $-598\pm 140 $ &$-11\pm 160$ & $608\pm 100$\vspace{6 pt} \\
    \bottomrule

    \end{tabular}
 \\
     \caption{Constraints on parameters for the monopole, dipole and quadrupole inferred in the Pantheon+ data set for different cuts in the redshift of the supernovae. For the sake of simplicity in presenting the results, we omit the constraints on dM. \label{tab:params_monopole_bulk_quadrupole}}
\end{table}

\subsection{Applying redshift cuts}
We repeat our \textit{bulk + monopole + quadrupole} analysis to sub-portions of the Pantheon+ dataset obtained by removing all the supernovae with a redshift smaller than a certain values $z_{\rm cut}$. Note, however, that all galaxies with Supernovae and Cepheids which are only used to determine $dM$ but do not affect the model, are always included, also if their redshift is below  $z_{\rm cut}$.
We have found that the inferred monopole, dipole and quadrupole is quite sensitive to the redshift cut for very low values of $z$ and becomes insensitive for $z_{\rm cut}>0.1$ For $z_{\rm cut}\geq0.05$ the monopole and dipole are actually no longer detected at more than 2$\si$. We therefore only report results for $z_{\rm cut}\leq 0.1$.
We show the resulting contour plots in figures~\ref{fig:contours_monopole_bulk_quadrupole_no_filter},~\ref{fig:contours_quadrupole_bulk_low_z} and in Appendix~\ref{appendix:extra_plots}, figures~\ref{fig:med} and~\ref{fig:high}. The constraints on monopole, dipole and quadrupole are presented in figure~\ref{fig:summary_constraints}.
In table~\ref{t1} we report the number of supernovae included within a given redshift cut.

\begin{table}[!ht]
\centering
\begin{tabular}{ c c c c }
\toprule
 z$_{\rm cut}$ & Pantheon+ & SNe in  \\
  & without Cepheids & Cepheid hosts & \\
 \midrule
 No cut & 1624 & 77 \\
 0.005 & 1615 & 50 \\  
 0.01 & 1576 &7  \\
 0.0175  & 1468 & 2  \\
  0.025  & 1312 &  0  \\
 0.0375  & 1126 & 0 \\ 
 0.05  & 1054 & 0 \\
  0.1  & 960 & 0  \\
 \bottomrule
 \end{tabular}
 \caption{Number of supernova lightcurves in each sub-portion of the Pantheon+ dataset obtained removing all the supernovae with a redshift smaller than $z_{\rm cut}$. (Note that while the Pantheon+ compilation contains 1550 different SNIa, it has 1701 lightcurves as several supernovae have been observed in more than one experiment.) For information, in the last column we provide also the number of SNe in Cepheid-host galaxies in each sub-dataset.  \label{t1} }
 \end{table}

\begin{figure}
\includegraphics [scale=0.33]{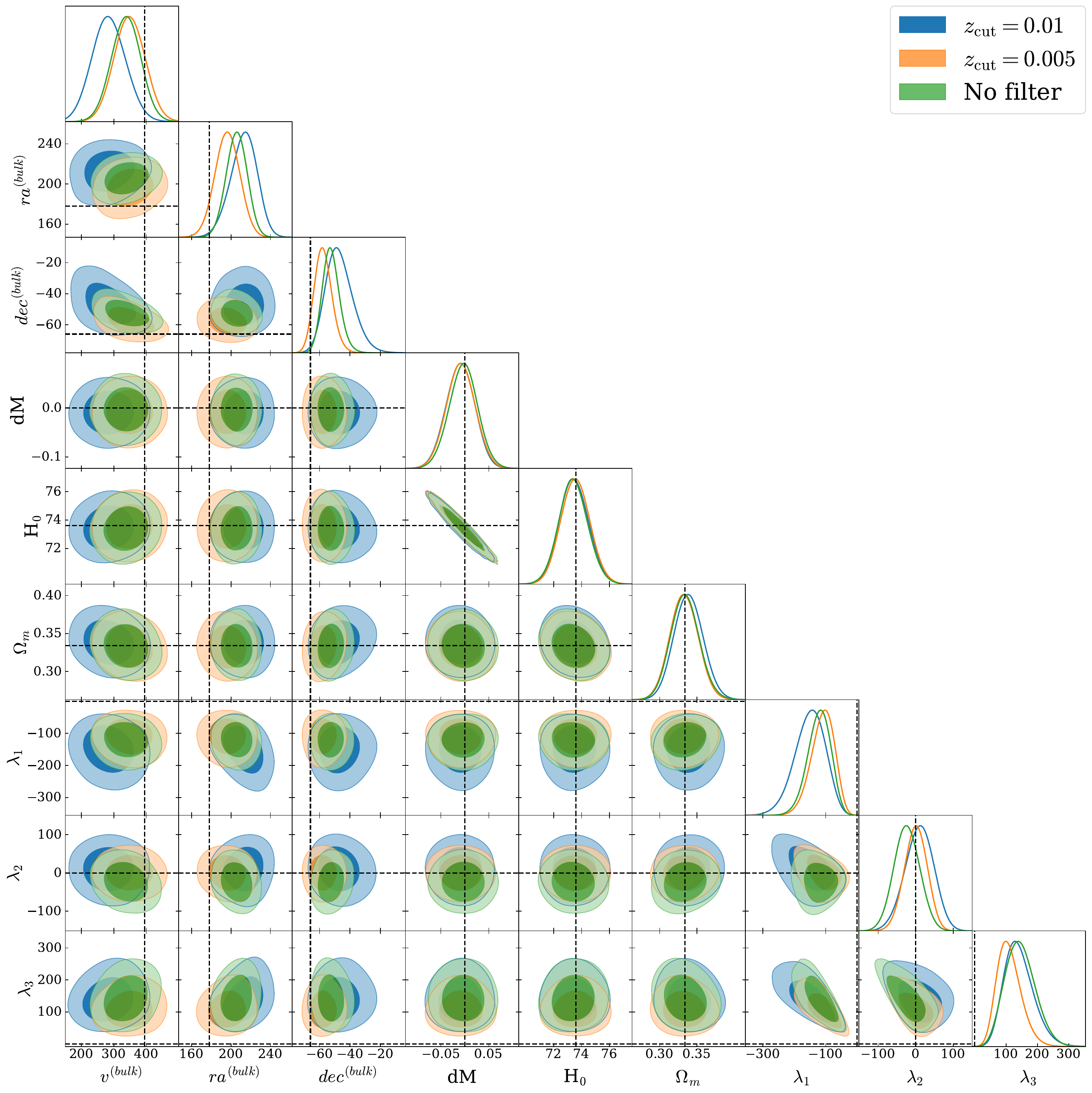}
	\caption{Contour plots for the bulk velocity and monopole analysis described in section~\ref{subsection:bulk_monopole_analysis} with three different cuts in the redshift of the supernovae. More cuts are shown in figures~\ref{fig:med} and~\ref{fig:high} in Appendix~\ref{appendix:extra_plots}.} \label{fig:contours_quadrupole_bulk_low_z}
\end{figure}

\begin{table}[!ht]
    \setlength\tabcolsep{5pt} 
    \scriptsize
    \begin{tabular}{cccccccccc}
        \toprule
		 $z_{\rm cut}$ &$|\bv^{(\rm bulk)}|$& ra$^{(\rm bulk)}$ & dec$^{(\rm bulk)}$&   $H_0$ & $\Omm$& $\lambda_1$  & $\lambda_2$& $\lambda_3$ \\ 
   		 &\scriptsize{[km/s]}&\scriptsize{[\textdegree]}&\scriptsize{[\textdegree]}&\scriptsize{[km/s/Mpc]}& &\scriptsize{[km/s]}&\scriptsize{[km/s]}&\scriptsize{[km/s]} \\
		\midrule
		No cut &  $338\pm 40 $ & $206\pm 11  $ & $-52.5^{+5.0}_{-6.0}$ & $73.4\pm 1.0$ & $0.334\pm 0.019$ & $-121^{+39}_{-33}$ &$-23\pm 34$ & $145^{+40}_{-50}$\vspace{6 pt} \\
		0.005 &  $349\pm 50 $ & $196\pm 12  $ & $-57.6^{+5.2}_{-6.2}$ & $73.6\pm 1.0 $ & $0.333\pm 0.019$ & $-110^{+41}_{-32}$ & $0\pm30$ & $110^{+30}_{-40} $ \vspace{6 pt} \\
		0.01 &   $432\pm 100$ & $242^{+30}_{-20}$ & $-65.0^{+7.6}_{-10}$ & $73.7\pm1.0$ & $0.334\pm 0.020$ & $-266^{+86}_{-71}$ & $-16\pm 70$& $281^{+80}_{-100}$\vspace{6 pt} \\
		0.0175 &  $363\pm 70 $ & $218^{+17}_{-14}$ & $-55.5^{+6.3}_{-8.2}$  & $73.5\pm 1.0$ & $0.340\pm 0.020$ & $-181^{+64}_{-50}$ & $16\pm46 $ & $165^{+40}_{-60}$\vspace{6 pt} \\
		0.025 &  $432\pm 100 $ & $242^{+30}_{-20} $ & $-65.0^{+7.6}_{-10}$  & $73.7\pm 1.0$ & $0.334\pm 0.020$ & $-266^{+86}_{-71}$ &  $-16\pm 70$& $281^{+80}_{-100}$\vspace{6 pt} \\
		0.0375 &  $316\pm200$ & $242^{+70}_{-30}$ & $-46^{+14}_{-32}$ & $73.5\pm 1.1$ & $0.339\pm 0.022$ & $-431^{+140}_{-120}$ & $28\pm 100$& $404\pm 100$\vspace{6 pt} \\
		0.05 & $210^{+100}_{-200}$ & $199^{+100}_{-80}$ & $-26^{+27}_{-50}$ & $73.8\pm 1.1$ & $0.328\pm 0.022$ & $-497^{+140}_{-120}$ & $5\pm 130$& $492 \pm 100 $\vspace{6 pt} \\
		0.1 & $280^{+100}_{-300}$ & $179\pm 100$ & $-10^{+39}_{-48}$  & $73.5\pm 1.1$& $0.334\pm 0.025$  & $-591\pm 140$ & $-12\pm 160$& $603\pm 100$ \vspace{6 pt} \\
		\bottomrule
    \end{tabular}
 \\
     \caption{Constraints on parameters for the dipole and quadrupole inferred in the Pantheon+ data set for different cuts in the redshift of the supernovae. For the sake of simplicity in presenting the results, we omit the constraints on dM. \label{tab:params_all_range_model_3}}
\end{table}

Our results are also summarized in table~\ref{tab:params_monopole_bulk_quadrupole}, where the full analysis is presented, and in table~\ref{tab:params_all_range_model_3} where we do not include the monopole. Note that already at $z=0.005$, the monopole is no longer detected at more than $1\si$. It is   significant only at very low redshifts where it is multiplied by a factor $1/H_0r(z)\simeq 1/z$. Nevertheless, its presence does somewhat raise the inferred value of the Hubble parameter, albeit within $1\si$ and not in the direction which would reduce the Hubble tension. Note also that for the first redshift cut only 9 supernovae in galaxies without Cepheids and 27 supernovae in galaxies with Cepheids are removed, see table~\ref{t1}. Hence most of the monopole signal comes from modelling  these 9 lowest redshift supernovae. 


This MCMC converges well also for $z_{\rm cut}=0.1$. The results for the bulk velocity and the quadrupole of this analysis are in good agreement with the full analysis. It is interesting to note that while for $z_{\rm cut}\geq 0.0375$ the dipole, i.e., the bulk velocity, is no longer detected at more than $1.5\si$, see table~\ref{tab:params_monopole_bulk_quadrupole} and
figures~\ref{fig:med_full_analysis} and~\ref{fig:high_full_analysis}, the eigenvalues $\lambda_1$ and $\lambda_3$ of the quadrupole remain non-zero even at 95\% confidence. Contrary to the bulk velocities, the eigenvalues $\lambda_1$ and $\lambda_3$ of the quadrupole are even increasing with redshift cut. This means that above $z_{\rm cut}=0.0375$, corresponding to a distance $R=112h^{-1}$Mpc, the angular fluctuations in the luminosity distance are better fitted with a quadrupole than with a dipole. This is not so surprising as the quadrupole represents fluctuating velocity field roughly on the scale of the redshift cut while the bulk velocity is assumed to be constant on all scales relevant in the analysis.

\subsection{ \texorpdfstring{$\chi^2$ analysis}{chi2 analysis} }
It is also interesting to study the improvement of the fit when including the monopole and quadrupole. As shown in table~\ref{tab:chi}, this strongly depends on the redshift cut.
Introducing no cut the fit is significantly improved and as we have seen this is mainly due to the monopole. At the very low cuts of $z_{\rm cut}=0.005$ and $z_{\rm cut}=0.015$ corresponding to a radius of $15 h^{-1}$Mpc and $30 h^{-1}$Mpc, the improvement is not significant considering that we have introduced 5 or, with monopole 6, additional parameters. However, for $z_{\rm cut}\geq 0.0175$, $\De\chi^2$ is monotonically increasing and for $z_{\rm cut}\geq 0.025$, the improvement obtained by including a quadrupole (and monopole which is irrelevant at this redshift) is, even if not overwhelming, more substantial. This is due to the fact that even for the highest redshift cuts, $\lambda_1$ and $\lambda_3$ are non-zero within 95\% confidence. On the contrary, their mean values are even increasing. Despite the fact that also the error bars increase, the significance of $\lambda_1$ and $\lambda_3$ simply measured as (mean value)/error also increases with redshift. This is not the case for $\bv^{\rm (bulk)}$ which can vanish within less than 90\% confidence for $z_{\rm cut}\geq 0.0375$ and becomes even less significant with increasing redshift, see table~\ref{tab:params_all_range_model_3} and figures~\ref{fig:med} and \ref{fig:high} in Appendix~\ref{appendix:extra_plots}. We finally note that whenever tested we found a slightly smaller $\chi^2$ when the correction is applied as a redshift correction, eq.~\eqref{theory:dl_z_quadrupole}, than when it is applied on the distance $
d_L$ directly, eq.~\eqref{theory:dl_quadrupole}.

\begin{table}[!ht]
    \centering
    \setlength\tabcolsep{20pt} 
    \begin{tabular}{c c c}
        & \multicolumn{2}{c}{$\Delta \chi^2$}  
        \vspace{3 pt} \\
        \toprule
        $z_{\rm cut}$ & \textit{Bulk + quadrupole} & \textit{Bulk + quadrupole} \\
         & & \textit{+ monopole} \\
        \midrule
        No cut & $5.92$ & $10.08$ \\
        $0.005$ & $3.15$ & $3.45$ \\
        $0.01$ & $6.40$ & $7.33$ \\
        $0.0175$ & $5.36$ & $7.07$ \\
        $0.025$ & $7.95$ & $8.06$ \\
        $0.0375$ & $8.04$ & $8.64$ \\
        $0.05$ & $9.37$ & $9.21$ \\
        $0.1$ & $9.52$ & $9.80$ \\
        \bottomrule
    \end{tabular}
    \caption{$\Delta \chi^2$ differences for different redshift cuts between the mean value dipole determined in our previous analysis~\cite{Sorrenti_2023} and the hypothesis of a bulk motion with only quadrupole correction (second column) and the hypothesis of a bulk motion with both quadrupole and monopole correction (third column).}
    \label{tab:chi}
\end{table}

\subsection{Mock tests \label{subsec:mock}}

\begin{figure}[!ht]
\centering
	\includegraphics [scale=0.6]{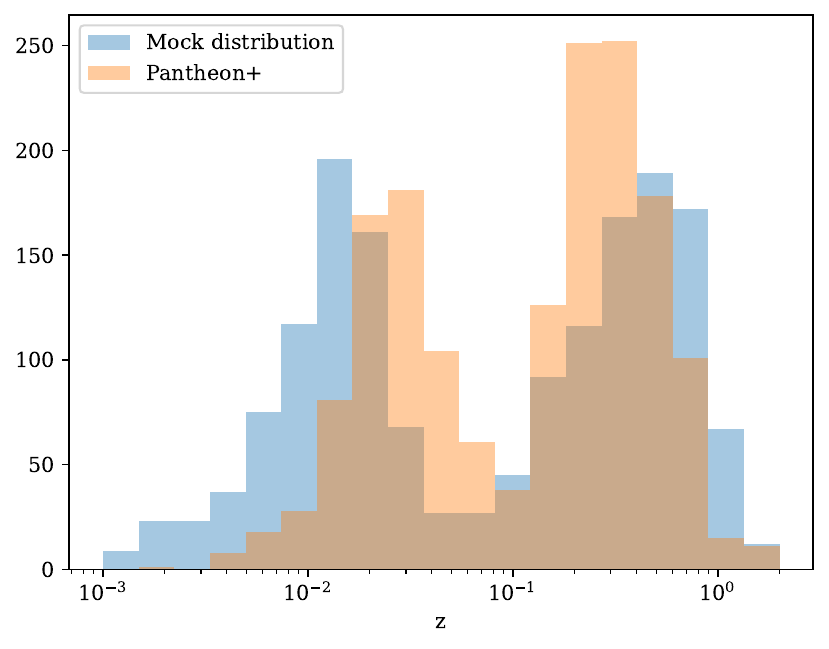}
	\caption{Visual comparison of the redshift distribution given by Pantheon+ (orange histogram) and the mock distribution (blue histogram) used in section~\ref{subsec:mock}.}
 \label{fig:mock_distribution}
\end{figure}

\begin{figure}[!ht]
	\includegraphics [scale=0.35]{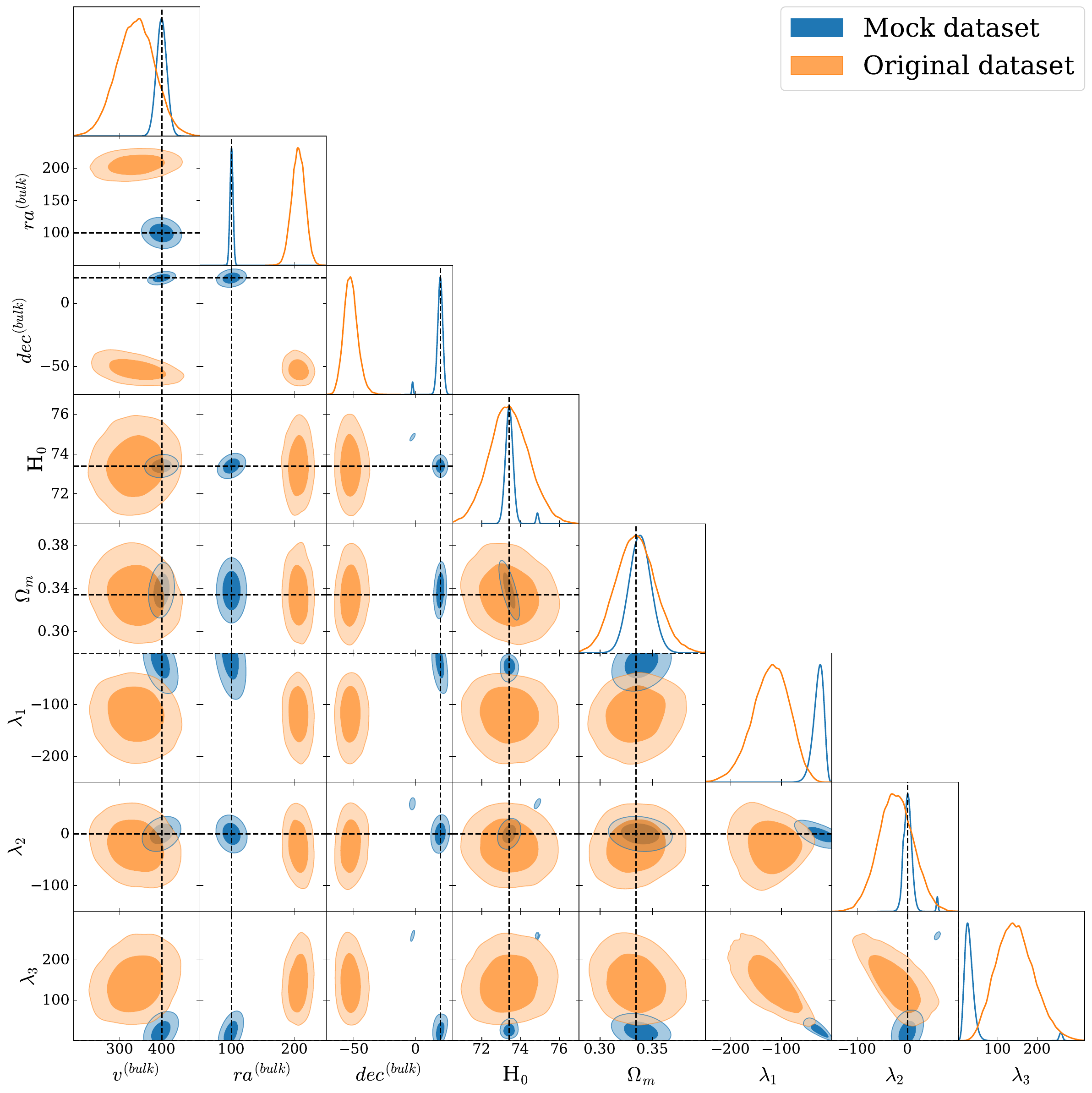}
	\caption{We compare the mean value parameters from the original Pantheon+ data (orange) with the ones from Mock data in which both, the monopole and the quadrupole have been removed. Note that $\la_1\leq\la_2\leq\la_3$ together with $\la_1+\la_2+\la_3=0$ enforces $\la_1\leq 0$ and $\la_3\geq 0$.} 
 \label{fig:mock_test}
\end{figure}

In order to confirm the significance of the monopole, quadrupole and the bulk  velocity (dipole in the supernova data), we have also compared the analysis of the true Pantheon+ data with the analysis of a mock dataset created using an artificial redshift distribution (the comparison with the original redshift distribution is shown in figure~\ref{fig:mock_distribution}), the distance moduli computed according to eqs.~(\ref{theory:z_quadrupole},~\ref{theory:v_p},~\ref{theory:dl_z_quadrupole}), fixing a constant $|\bv^{(\rm bulk)}|=400$km/s in direction $({\rm ra,dec}) =(100$\textdegree, $20$\textdegree), so that the monopole and quadrupole vanish, and a cosmology given by the fiducial values $\Omm=0.334$, $H_0=73.6$. Moreover, for simplicity, we neglected supernovae hosted in cepheids, fixing dM=0, and we used a diagonal covariance whose elements are the same as the diagonal of the covariance matrix of the Pantheon+ analysis. 

For $H_0$, $\Omm$ and the bulk velocity we are able to recover the values chosen for the construction of the mock dataset. At the same time, we do not find any monopole or quadrupole, while they are detected in the real data. In figure~\ref{fig:mock_test} we show the results from a parameter estimation of the Mock dataset  compared to the real data for dipole and quadrupole only. The bulk velocity inserted in the Mock data and the cosmological parameters used are indicated by dashed vertical and horizontal lines. Clearly, the input parameters are very well reproduced and all three eigenvalues of the quadrupole are well consistent with zero. The same is true for the monopole. This confirms our interpretation that the bulk velocity, the monopole and the quadrupole are really present in the data.

\section{Systematics checks}
\begin{figure}[!ht]
\centering
	\includegraphics [scale=0.8]{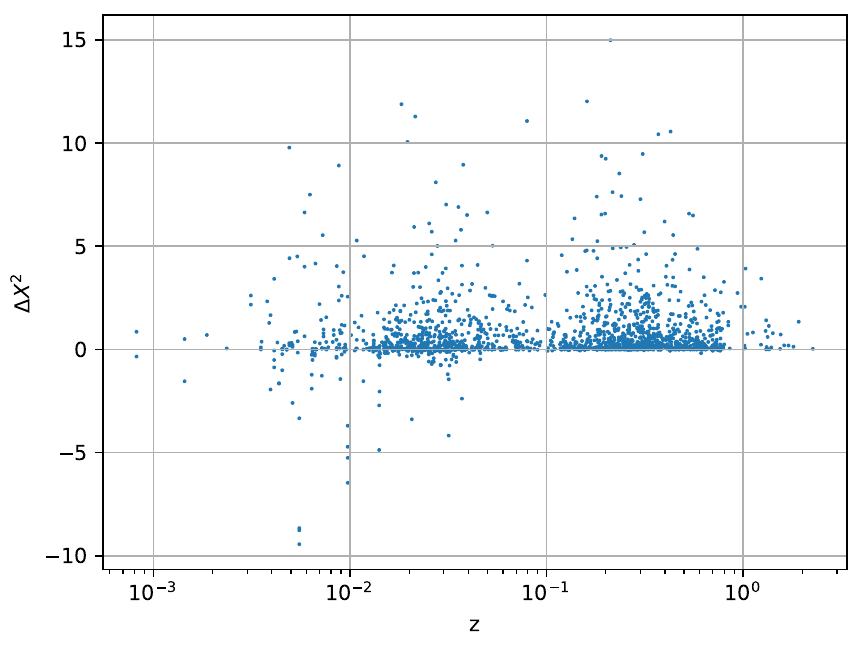}
\caption{$\chi^2$ differences as function of redshift for the full \textit{bulk + monopole + quadrupole} analysis between considering all the supernovae and removing one supernova at a time.}
\label{fig:chi_test}
\end{figure}

We tested wether our signal was dominated by a single supernova. For doing so, we first computed the $\chi^2$ differences for the full \textit{bulk + monopole + quadrupole} analysis between considering all the supernovae and removing one supernova at a time. We plot the differences as function of the redshift in fig.~\ref{fig:chi_test}. We then run an MCMC removing the 4 supernovae with redshift $z\leq0.1$ contributing the most to $\chi^2$ (i.e. the supernovae whose $\Delta \chi^2\geq 10$). We considered only $z<0.1$ supernovae since they are the ones mainly constraining the multipoles. 
From the MCMC run, we obtained essentially identical contours as in fig.~\ref{fig:contours_monopole_bulk_quadrupole_no_filter}, proving that the our signal is not dominated by a one or a few supernovae.

\begin{figure}[!ht]
\centering
	\includegraphics [scale=0.8]{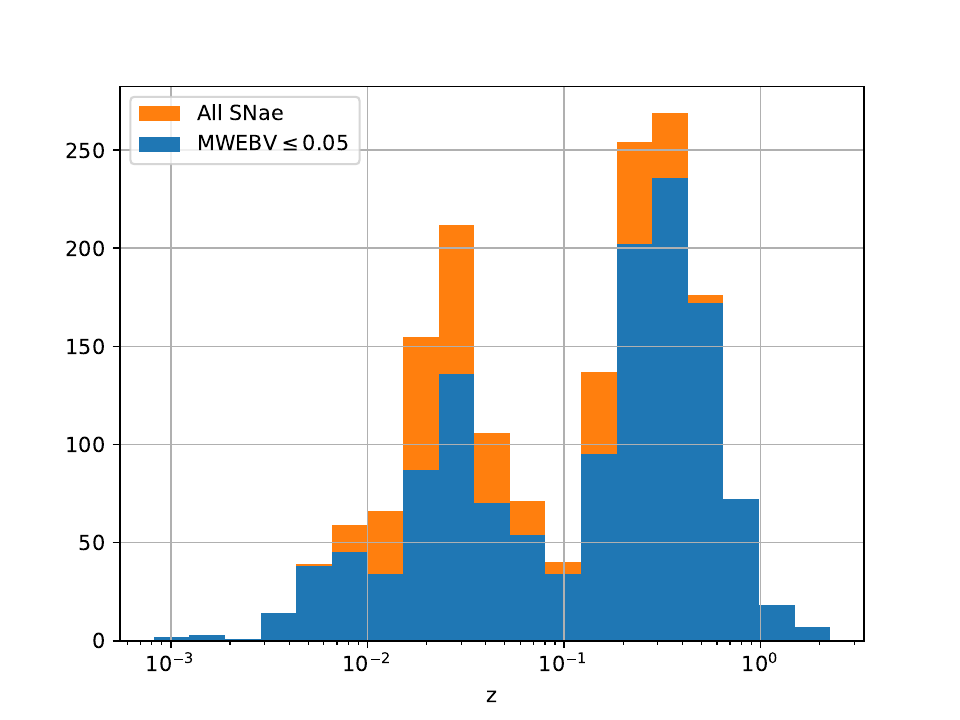}
\caption{Comparison of the full supernovae distribution (orange histogram) with the distribution of the 1320 supernovae with MWEBV$\leq0.05$ (blue histogram).}
 \label{fig:hist_mwebv}
\end{figure}

\begin{figure}[!ht]
\centering
	\includegraphics [scale=0.3]{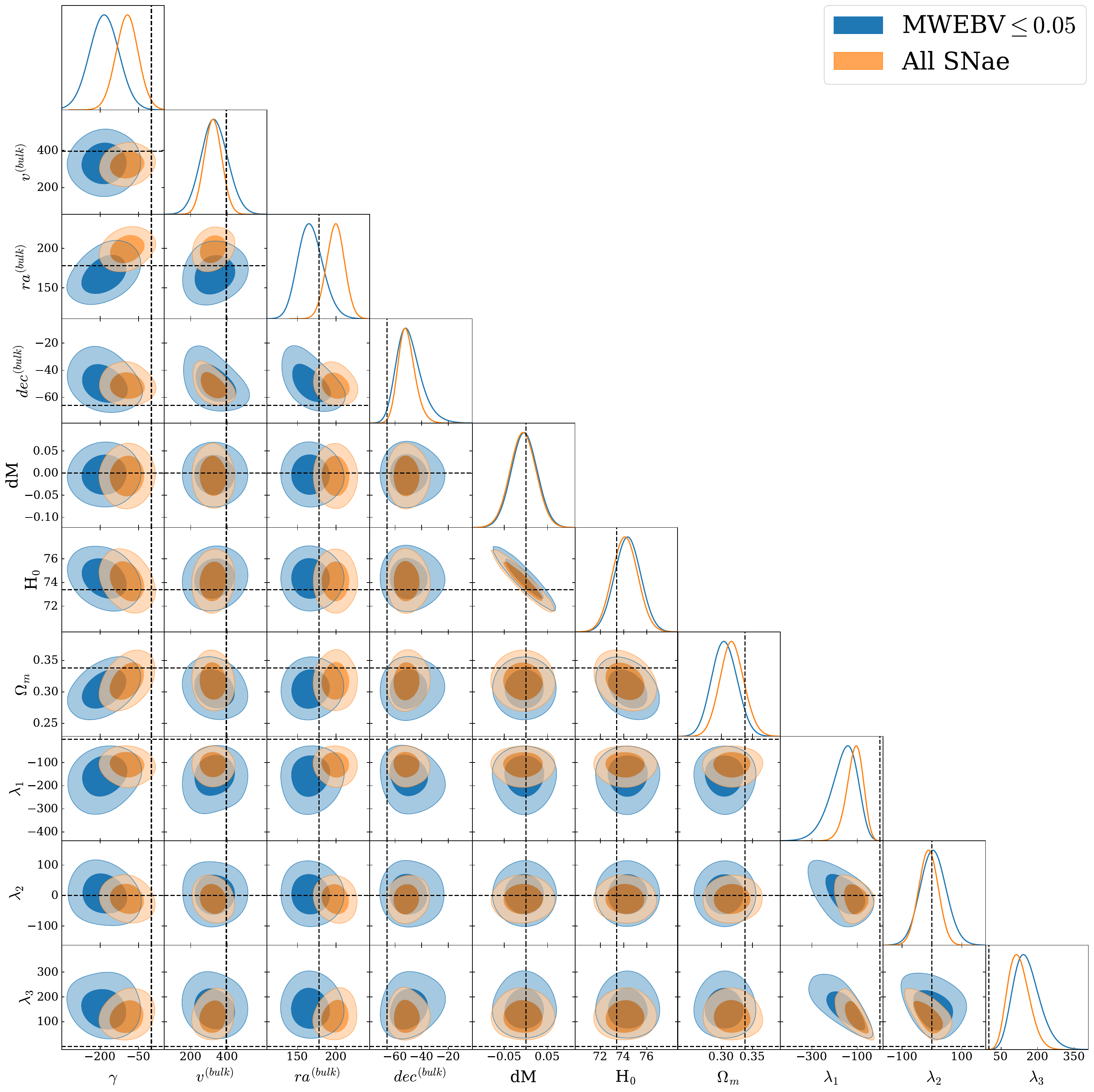}
    \caption{Contour plots for the full bulk velocity, quadrupole and monopole analysis described in section~\ref{subsection:full_analysis} considering all the supernovae (orange contours) and the 1320 supernovae with MWEBV$\leq0.05$ (blue contours).}
 \label{fig:mwebv_multipoles}
\end{figure}

We also checked wether our signal was the result of a Milky Way dust systematic in the supernovae. For doing so, we performed our \textit{bulk + monopole + quadrupole} analysis on a subdataset of 1320 supernovae with the Milky Way extinction parameter MWEBV$\leq0.05$ (see fig.~\ref{fig:hist_mwebv}). We obtained the contours in fig.~\ref{fig:mwebv_multipoles}. The contours are of course somewhat wider due to the lower statistics, we removed 381 supernovae from our dataset, but the main results remain valid. There is a shift for the monopole and the right ascension of the bulk velocity at the level of nearly 1$\sigma$. The difference between the mean value for the bulk velocity obtained in the main analysis in section~\ref{subsection:full_analysis} and the one obtained when removing supernovae with MWEBV$\leq0.05$ is
\be
\Delta v^{\text{(bulk)}}=7{\rm km/s} \,, \qquad  
({\rm \Delta ra, \Delta dec}) =(32, -1.7)\,,
\label{e:DvMWEBV}
\ee
while the difference between the monopole mean values is
\be
\Delta \gamma = -89 {\rm km/s}
\ee
While the monopole becomes even more negative and actually more significant, the trend for the bulk velocity is less clear. We therefore just add the difference due to Milky Way dust extinction
as systematic uncertainties in quadrature to the errors inferred from the Pantheon+ covariance matrix that is given in table~\ref{tab:params_monopole_bulk_quadrupole}. The resulting errors for the bulk velocity are shown in table~\ref{t:MWEBV}.
\begin{table}
\begin{center}
\begin{tabular}{ccc}
\hline
$v^{(\text{bulk})}$[km/sec] & ra $[^o]$ &  dec $[^o]$\\ \hline
$323 \pm 50.5$ & $199 \pm 34$ & $-51.2^{+5.4}_{-7.2}$ \\
\hline
\end{tabular}
\end{center}
\caption{The bulk velocity when adding a systematic error due to Milky Way dust extinction. \label{t:MWEBV}}
\end{table}

Clearly, adding these possible errors does not  change our main results. The bulk velocity remains significant and even though the right ascension is now within $1\si$ of the observer velocity $v_\odot$ given in \eqref{e:d-Planck}, the declination is still very far away, it rather becomes even more negative.

\section{Conclusions}
In this paper we analysed the Pantheon+ data including a dipole, a quadrupole and a monopole perturbation in the luminosity distance which are motivated by the peculiar motion of the supernovae which leads to an angular dependence of the redshift perturbation. We have found that both the quadrupole and at very low redshift also  the monopole are significant and of a comparable amplitude as the dipole of the bulk velocity. Removing low redshift supernovae from our sample, we even find that the monopole and the dipole from the bulk motion are no longer detected with high significance. This can be due to the fact that at higher redshifts, monopole and dipole perturbations, which are sensitive to fluctuations that are at least of the size of the redshift shell, would require fluctuations on ever larger scales which, within standard cosmology are small. However, the quadrupole, that measures fluctuations of typically half the size of the shell remains non-zero at more than 95\% confidence. It is also interesting to note that the eigenvalues of the quadrupole are significantly increasing with redshift while their errors stay roughly constant. Hence they also become more significant at higher redshift. 
This trend is understandable as at higher redshifts, where neither the monopole perturbation nor the dipole (bulk velocity) are significant, the  peculiar velocity of the sources has to be modelled by the quadrupole alone in our approach. 

\begin{figure}[!ht]
	\includegraphics [scale=0.47]{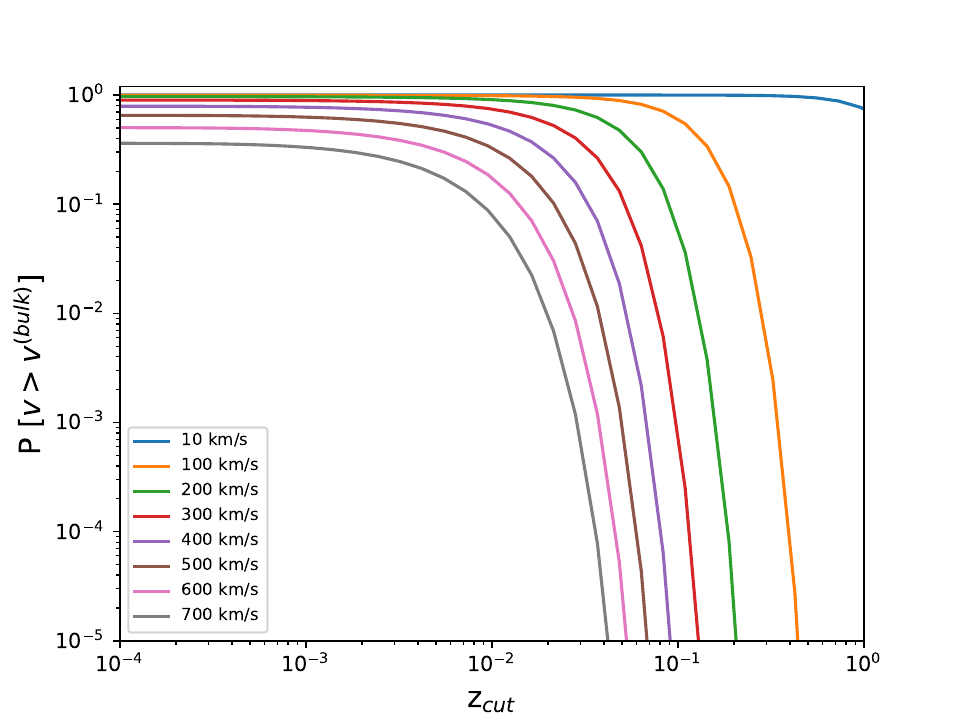}
 \includegraphics [scale=0.47]{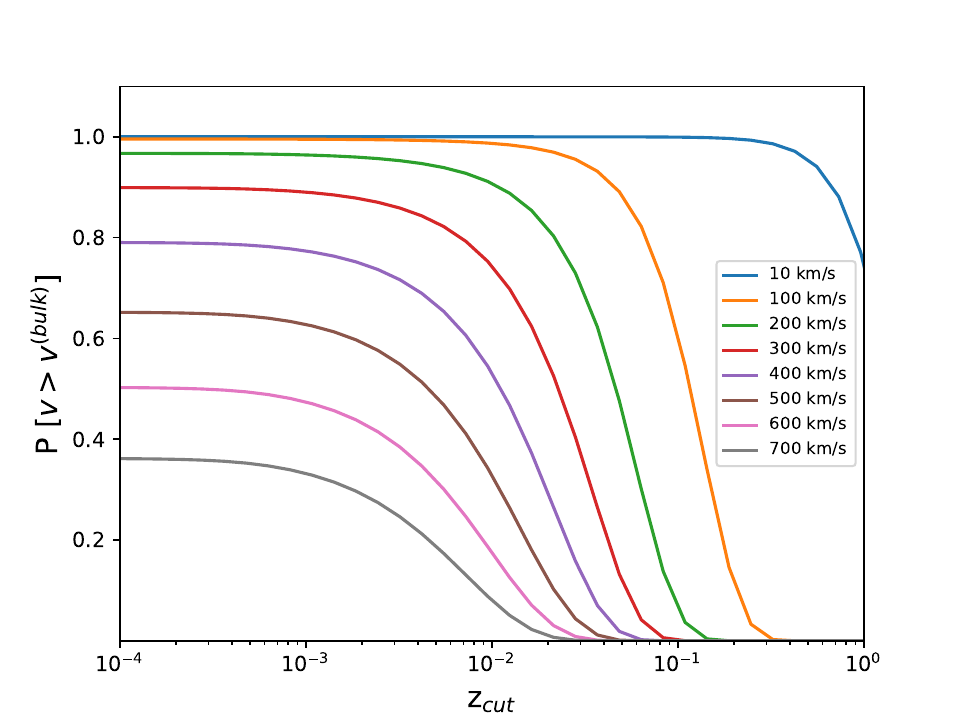}
	\caption{We show the probability to measure a bulk velocity larger than a given value inside a ball corresponding to the redshift $z_{\rm cut}$. For better visualisation we show the result in  log scale (left panel) and linear scale (right panel).}
 \label{fig:proba}
\end{figure}

It is intriguing that our result for the bulk velocity is within 95\% confidence in agreement with the bulk flow found in the CosmicFlows4 analysis~\cite{watkins23}. That it does not perfectly agree with CosmicFlows4 is not surprising since our $z_{\rm cut}$ means that we exclude all supernovae inside a ball of radius $r(z_{\rm cut})$. While the remaining bulk flow is dominated by the one in a shell close to $r(z_{\rm cut})$, this is not quite the same as the velocity inside the ball. Also, while the CosmicFlows4 analysis does include the Pantheon+ data, it has a much larger catalog of about 38,000 galaxy velocities. For $z_{\rm cut}<0.0375$ we obtain a bulk flow of about $316\pm200$km/s, and based on Appendix~\ref{app:stat}, see also figure~\ref{fig:proba}, we conclude that the probability to find a velocity of this size or larger inside a ball of redshift $z_{\rm cut}=0.0375$ is
\be
P(v^{\rm(bulk)}\geq 316,z_{\rm cut}=0.0375) = 16\% \,, 
\ee
where we have used the cosmological parameters inferred from the MCMC analysis reported in table~\ref{tab:params_monopole_bulk_quadrupole}. This is the result within standard $\La$CDM with $\si_{kV}(r=112h^{-1}$Mpc$) \simeq 212$km/s, and arguably does not appear to be highly unlikely. Taking additionally into account that our error in $v^{\rm(bulk)}$ is relatively large, our analysis does not exclude $v^{\rm(bulk)}=(316-200)$km/s, for which we find
\be
P(v^{\rm(bulk)}\geq 116,z_{\rm cut}<0.0375) = 93.73\% \,, 
\ee
and for which certainly there is no tension. 

The reason that our results are not in strong tension with $\La$CDM is mainly due to the weaker statistical power of the supernova-only sample, hence to the large error bars of $v^{\rm(bulk)}$ and to the fact that we have no truly significant bulk flow at $r\geq 200h^{-1}$Mpc. Were we to take at face value the bulk flow of 274km/s at $z_{\rm cut}=0.1$ reported in table~\ref{tab:params_monopole_bulk_quadrupole}, which corresponds to $r=300h^{-1}$Mpc,
with $\si_{kV}(r=300h^{-1}$Mpc$) \simeq 98.69$km/s, we would find
\be
P(v^{\rm(bulk)}\geq 274,z_{\rm cut}<0.1) = 3.82\times 10^{-5} \,. 
\ee
This would be similar to the findings of Ref.~\cite{watkins23}. But from table~\ref{tab:params_monopole_bulk_quadrupole} we also see that $v^{\rm(bulk)}(z=0.1)$ is compatible with zero within 2$\sigma$, and is statistically acceptable for the $1\sigma$ lower limit.

Within the statistical power of the Pantheon+ data, we therefore conclude that the inferred monopole perturbation, dipole and quadrupole in the Pantheon+ data are in reasonable agreement with a velocity field expected in the standard $\La$CDM model of cosmology. It will certainly be very important to repeat this analysis  with a larger sample of supernovae. Especially low redshift supernovae with $z\leq 0.1$ are suited to improve the statistical power. For example, the Vera Rubin Observatory's LSST should be able to detect up to $10^4$ supernovae within a year or so in this redshift range~\cite{LSSTDarkEnergyScience:2024nox}, while the Pantheon+ dataset has 800 sources in this redshift range.

\acknowledgments
We thank Eoin Ó Colgáin and Subir Sarkar for interesting comments.
The authors acknowledge financial support from the Swiss National Science Foundation. The computations were performed at University of Geneva using {\it Baobab} HPC service.

\vspace{2cm}

\appendix

\section{Pantheon+ redshift dependence } \label{appendix:extra_plots}
\begin{figure}
	\includegraphics [scale=0.33]{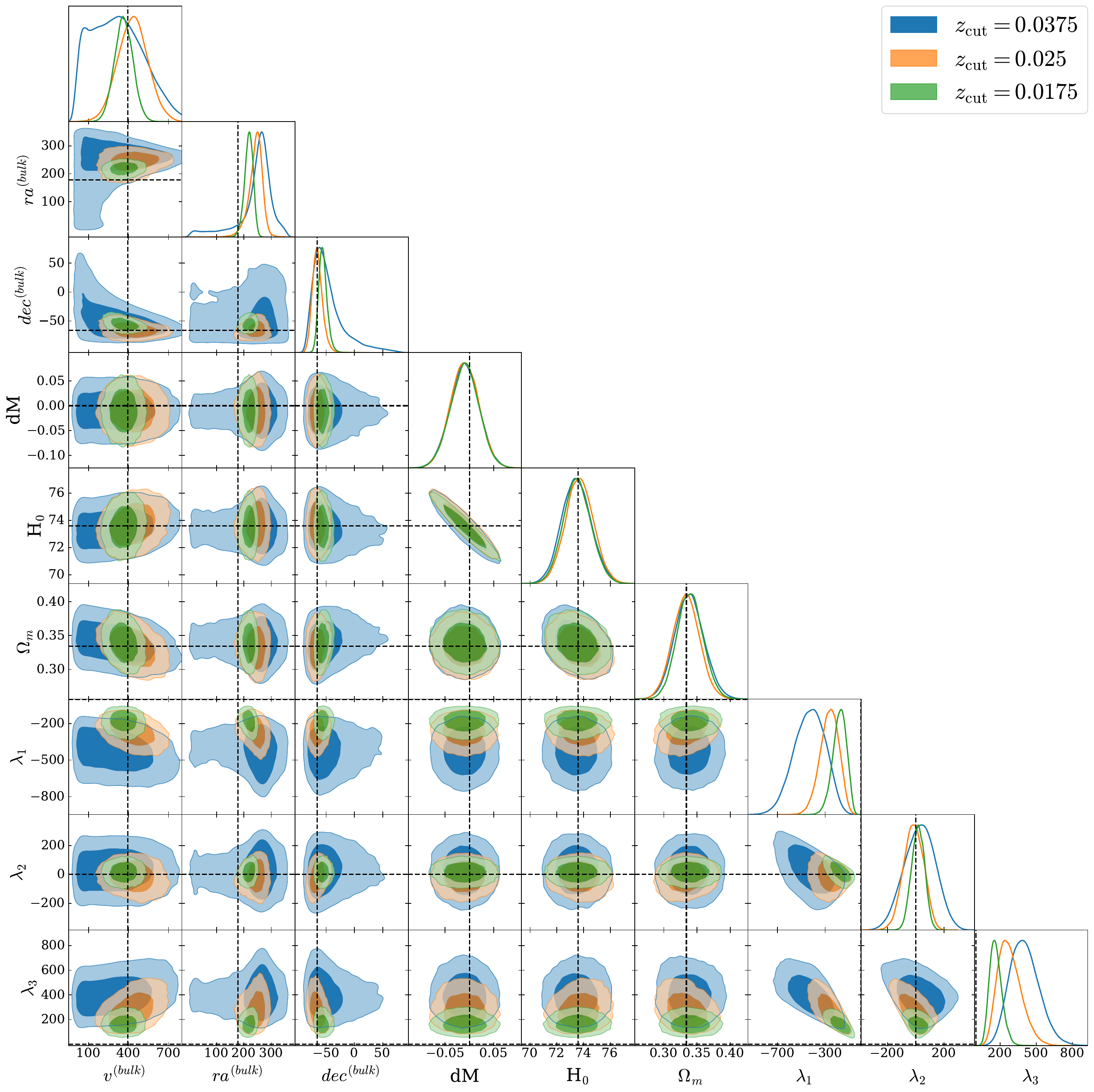}
	\caption{Parameters for redshift cuts at medium redshifts for the \textit{bulk + quadrupole} analysis.}
 \label{fig:med}
\end{figure}

\begin{figure}

	\includegraphics [scale=0.33]{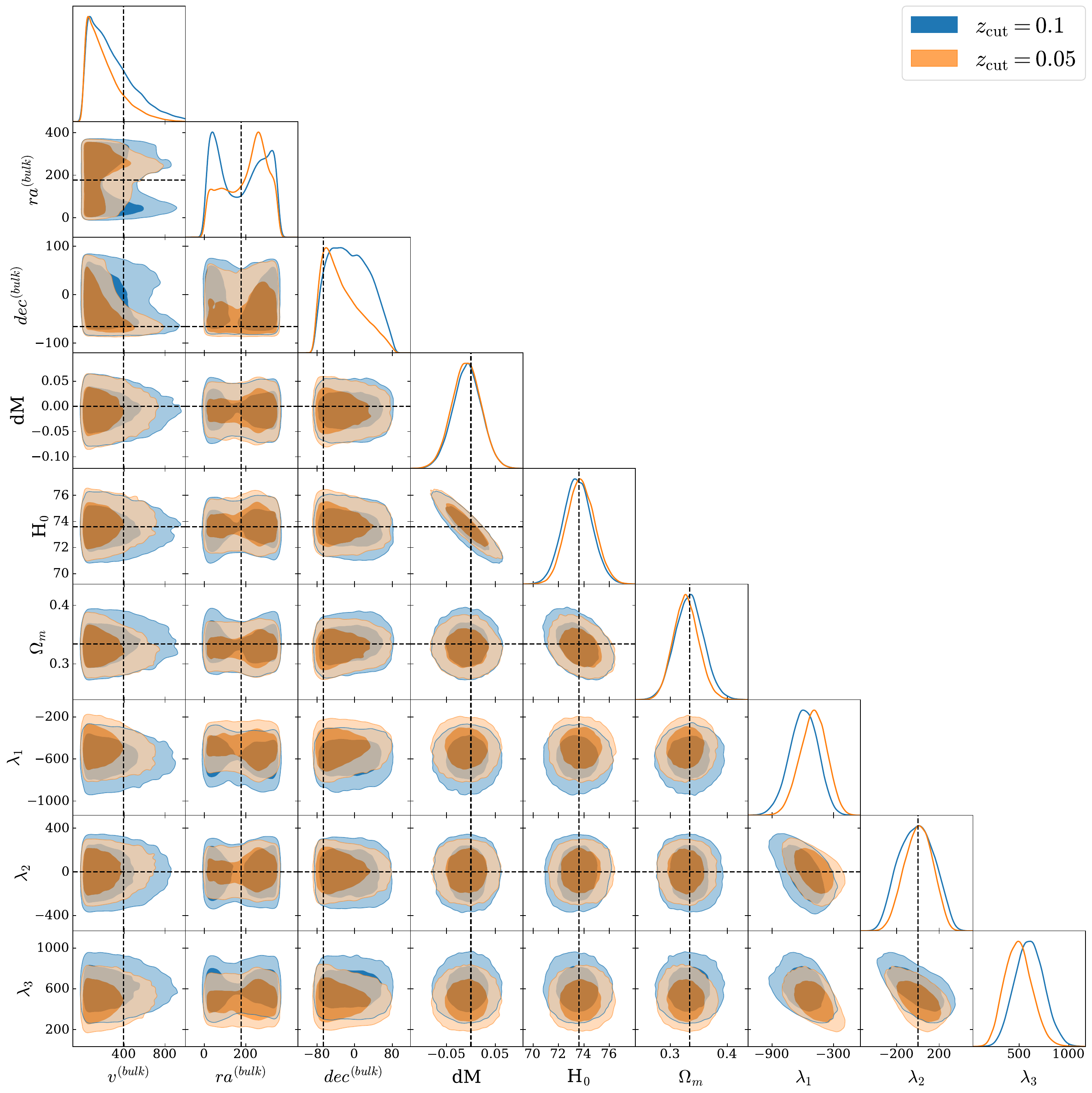}
	\caption{Parameters for redshift cuts at relatively high redshifts for the \textit{bulk + quadrupole} analysis.}
 \label{fig:high}
\end{figure}

\begin{figure}
\centering
	\includegraphics [scale=0.30]{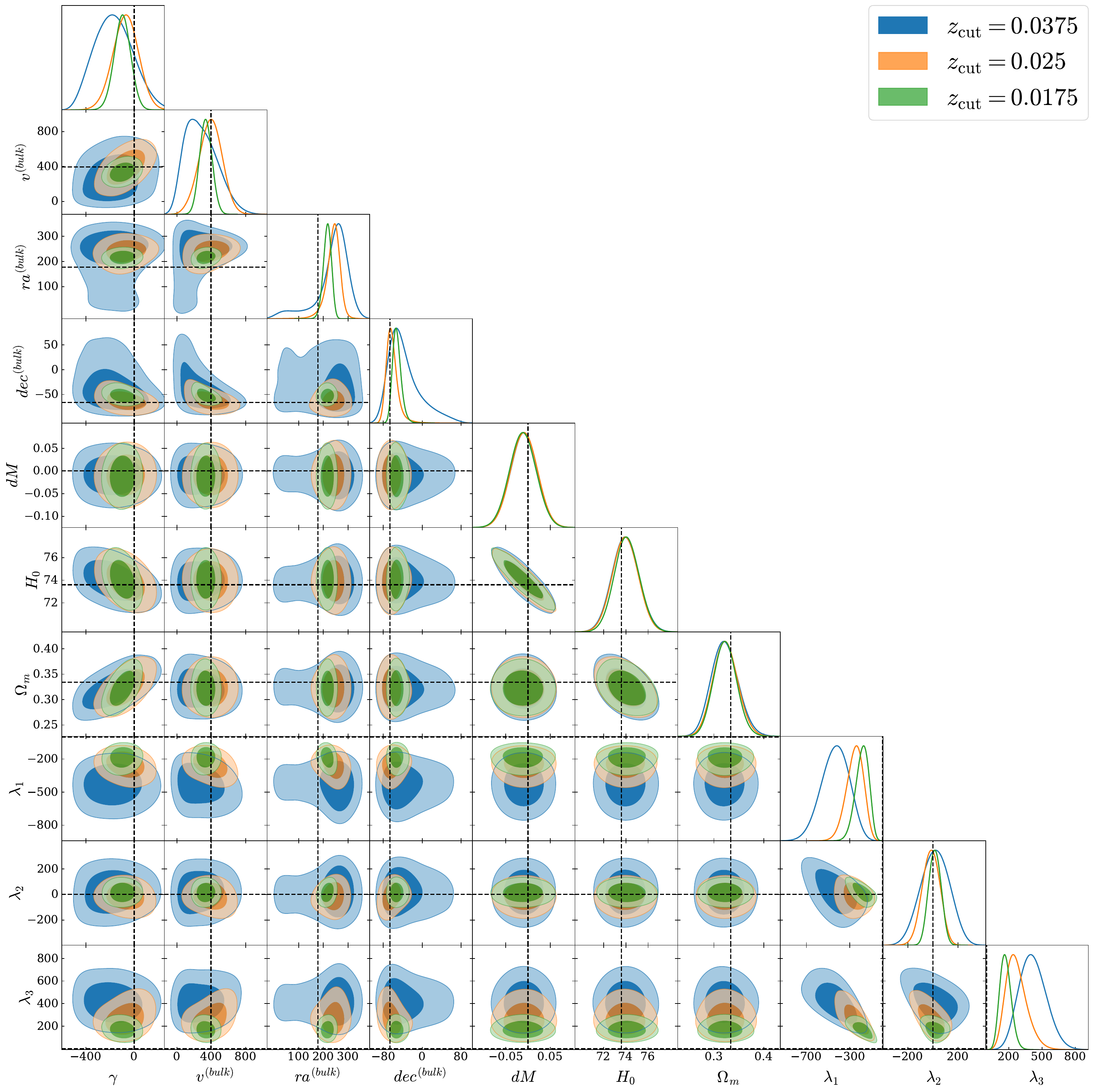}
	\caption{Parameters for redshift cuts at medium redshifts for the full \textit{bulk + quadrupole + monopole} analysis.}
 \label{fig:med_full_analysis}
\end{figure}

In this appendix we also show the contour plots for higher  redshift cuts for both the \textit{bulk + quadrupole} analysis and the full \textit{bulk + quadrupole + monopole} analysis. We choose $z_{\rm cut}= 0.0175,~0.025$ and $0.0375$ in figure~(\ref{fig:med}, \ref{fig:med_full_analysis}) for both the analyses, whereas we set $z_{\rm cut}= 0.05$ and $0.1$ in figure~\ref{fig:high}, for only the \textit{bulk + quadrupole} analysis.

\begin{figure}
\centering
	\includegraphics [scale=0.30]{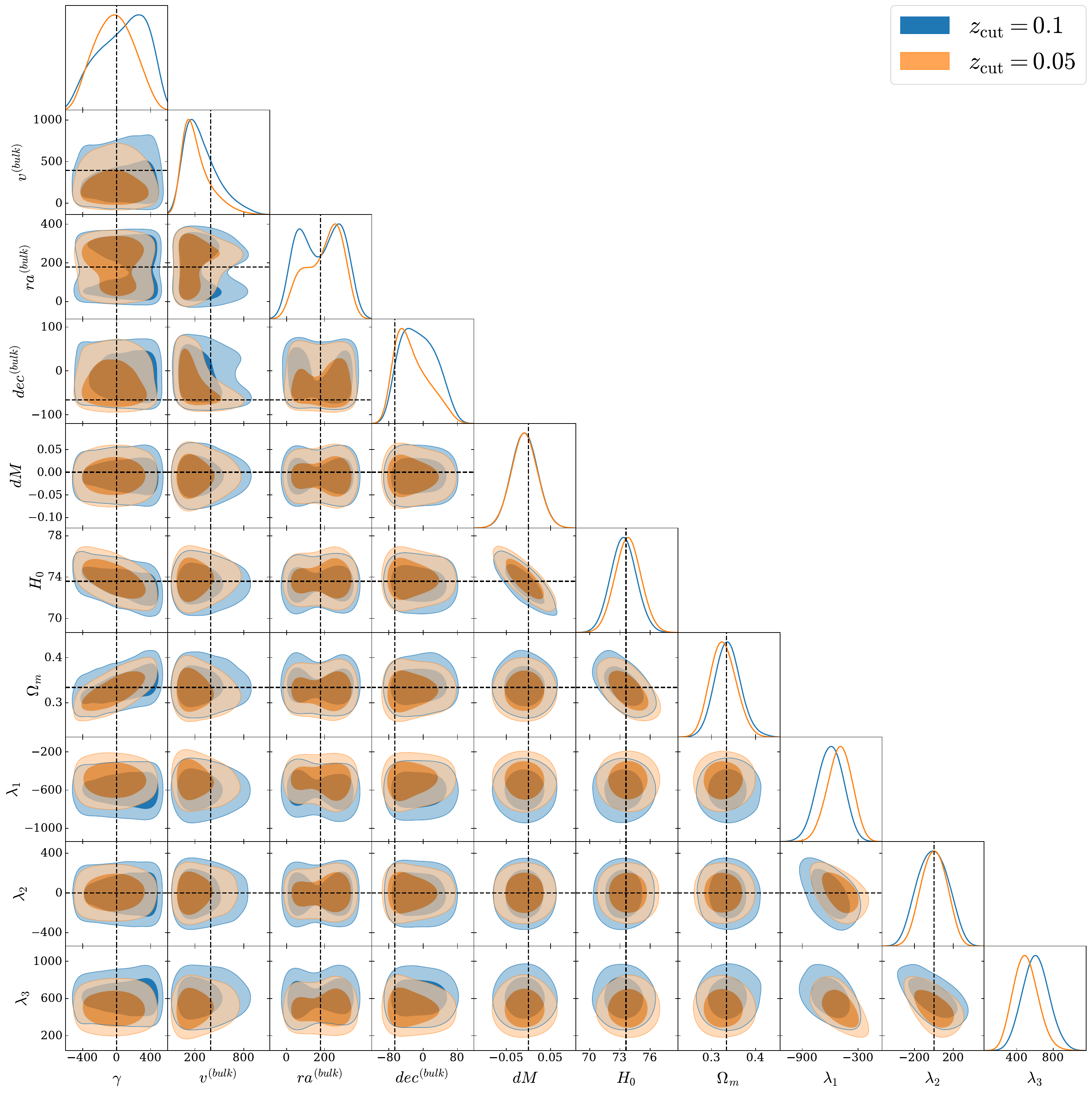}
	\caption{Parameters for redshift cuts at relatively high redshifts for the full \textit{bulk + quadrupole + monopole} analysis.}
 \label{fig:high_full_analysis}
\end{figure}

\FloatBarrier

\section{Statistical properties of the velocity field}
\label{app:stat}
In the standard cosmological model, the velocity field $\bv$ is the gradient of a scalar velocity potential $V$, $\bv = \nabla V$. The velocity potential is an isotropic Gaussian random field with zero mean and power spectrum (at $z=0$) $P_V(k)$. This implies that each component $v_j$ of the velocity field is itself an isotropic Gaussian random field, and in Fourier space is given by
\be
v_j = i k_j V \, .
\ee
The variance of the fields $v_j$ and $V$ are then related through
\be
\langle v_j^2 \rangle = k_j^2\langle  V^2 \rangle \, .
\ee
Thanks to the statistical isotropy of the velocity field, each component $k_j^2$ contributes equally to $k^2=k_1^2+k_2^2+k_3^2$, so that the prefactor is $1/3$ on average, and
\be
\sigma^2_{v_j} = \frac{k^2}{3} \sigma^2_{V} =\frac{1}{3} \sigma^2_{kV}\, .
\ee

\begin{figure}
    \centering
    \includegraphics[width=0.65\linewidth]{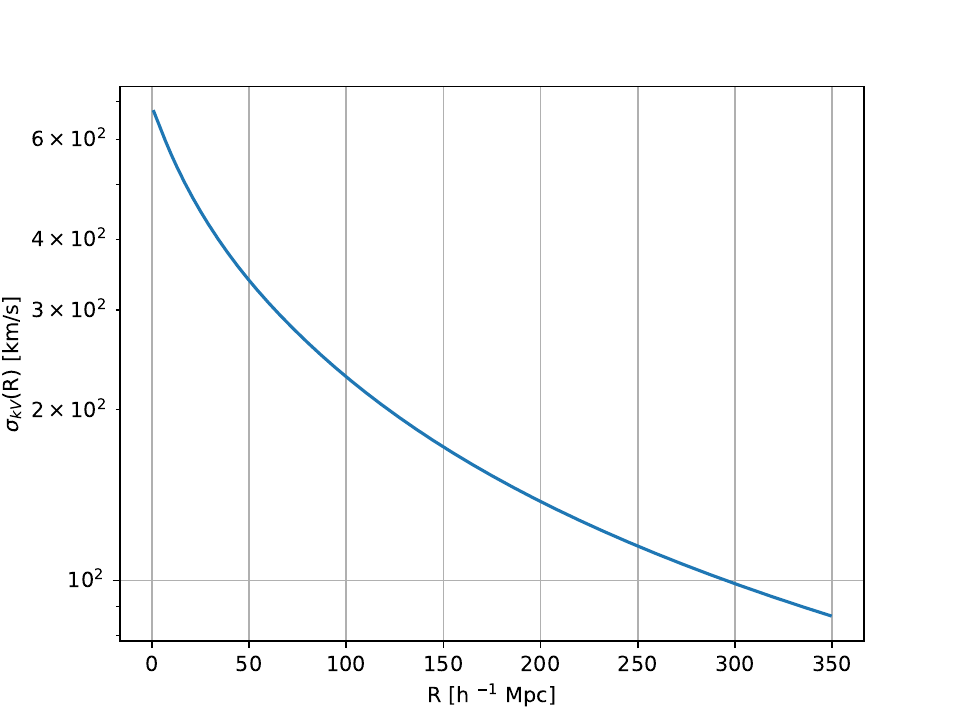}
    \caption{The variance of the velocity  $kV$, averaged over a spherical top-hat with radius $R$.}
    \label{fig:sigmav}
\end{figure}

Here we are interested in the variance of the velocity field when averaged over a spatial volume of a given size $R$,
\be
\sigma^2_{kV}(R) = \int P_{kV}(k) | W_R^2(k)| \frac{d^3\! k}{(2\pi)^3} 
\ee
for a given window function $W_R(k)$ that describes the shape of the spatial volume. For a spherical top-hat window in real space, the Fourier-space window function is
\be
W_R(k) = \frac{3}{(kR)^3} \left(\sin(k R) - k R \cos(kR) \right) = \frac{3}{k R} j_1 (k R) \, .
\ee
In perturbation theory, the power spectrum of $V$ is related to the power spectrum of the density contrast $\delta$ through
\be
P_{kV}(k) = H_0^2 f_0^2 \frac{P_\delta(k)}{k^2}
\ee
where $f_0 = f(z=0) $ is the growth factor today in the $\Lambda$CDM model. We show $\sigma_{kV}(R)$ as a function of $R$ in figure~\ref{fig:sigmav}.\footnote{For computing the power spectrum used in the probability analysis we used the boltzmann solver \texttt{camb}~\cite{camb}, assuming $H_0$=73.6 km/s/Mpc and $\Omm$=0.334 for the background cosmology.} 

\begin{figure}
    \centering
    \includegraphics[width=0.65\linewidth]{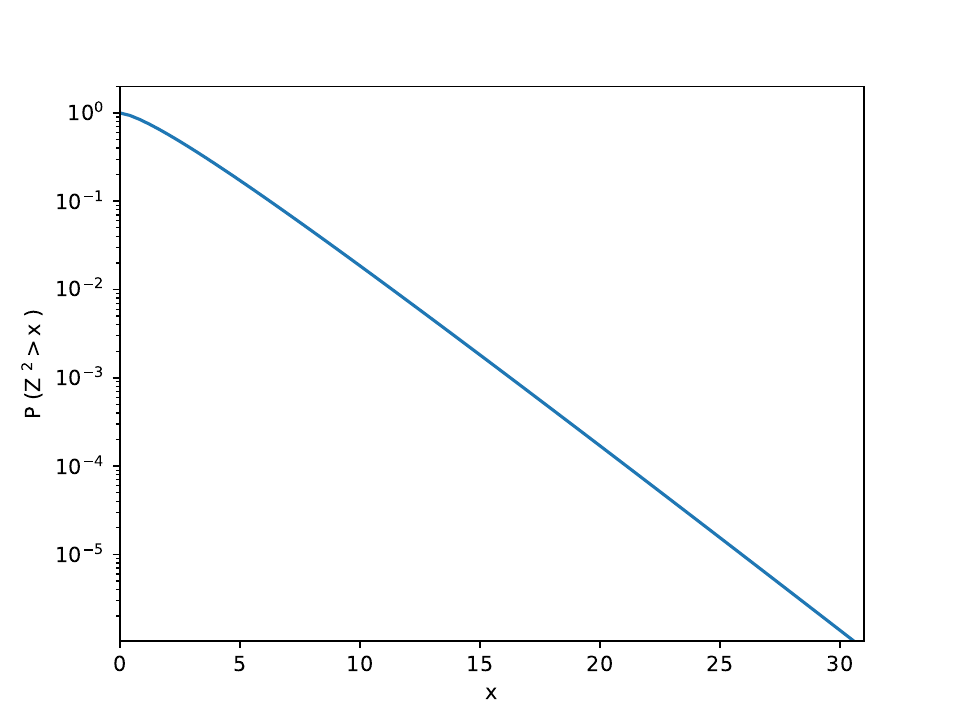}
    \caption{The probability to find a value large than $x$ for a random variable with a $\chi^2$ distribution for 3 degrees of freedom.}
    \label{fig:chisqcum}
\end{figure}

In order to evaluate the probability for finding a larger bulk velocity on a given scale $R$ than a certain value $v_0$ we consider the random variable
\be
Z_j = \frac{\sqrt{3} \, v_j}{\sigma_{kV}(R)}\,,
\ee
which has zero mean and unit variance. Its norm-squared, $Z^2 = \sum_j Z_j^2 = 3 v^2/\sigma^2_{kV}(R)$ then has a $\chi^2$ distribution with 3 degrees of freedom (and $|Z|$ has a $\chi$ distribution). The probability to find a value larger than $x$ for a random variable $Z^2$ that has a $\chi^2$ distribution with $n$ degrees of freedom is
\be
P(Z^2 > x) = \frac{\Gamma(n/2,x/2)}{\Gamma(n/2)}
\ee
where $\Gamma(k,x)$ is the incomplete Gamma function. We show this probability for $n = 3$ in figure~\ref{fig:chisqcum}.
From figure~\ref{fig:sigmav} we see that on a scale of $R=112h^{-1}$Mpc which corresponds to $z_{\rm cut}=0.0375$, we find $\sigma_{kV} \simeq 212$ km/s. In our analysis we find a mean value of $v^{\rm (bulk)} \approx 374$ km/s inside this redshift cut. For a $\chi^2(3)$ distribution, we obtain
\be
P(Z^2 > 3(374/212)^2) \approx 0.01599 \, .
\ee 
While this probability is not very high, it is by no means excluding the model. 

\bibliography{refs}
\bibliographystyle{JHEP}

\end{document}